\documentclass[fleqn,usenatbib]{mnras}

\usepackage[T1]{fontenc}
\usepackage{ae,aecompl}

\usepackage{graphicx}	
\usepackage{amsmath}	
\usepackage{multicol}   
\usepackage{bm}		    
\usepackage{pdflscape}	
\usepackage{epsfig,color,pifont}
\usepackage{times}
\usepackage{subfig}
\usepackage{url}

\usepackage{graphicx}	
\usepackage{amsmath}	
\usepackage{amssymb}	

\usepackage{newtxtext,newtxmath}

\usepackage{physics}  

\def\Msun{\mbox{~M$_\odot$}}

\def\Msunpc2{\mbox{~M$_\odot$~pc$^{-2}$}}

\def\kms{\mbox{~km~s$^{-1}$}}

\def\kpc{\mbox{~kpc}}
\def\Mpc{\mbox{~Mpc}}

\def\R200{R_{200}}
\def\M200{M_{200}}

\def\Mstar{M_*}

\def\Mhalo{M_{\rm halo}}

\def\Mstargal{\Mstar^{\rm gal}}
\def\Re{R_{\rm e}}
\def\Regal{\Re^{\rm gal}}
\def\feh{\rm [Fe/H]}
\def\fehgal{\feh_{gal}}
\def\fehgc{\feh}
\def\alphagc{\rm [\alpha/Fe]}
\def\alphagal{\rm [\alpha/Fe]_{gal}}
\def\deltafeh{\Delta \feh}
\def\deltaalpha{\Delta \alphagc}
\def\sigmagal{\sigma_{\rm gal}}
\def\sigmagcs{\sigma_{\rm GCs}}
\def\Rp{R_{\rm p}}
\def\Vp{V_{\rm p}}
\def\Vrot{V_{\rm rot}}

\def\Ngc{N_{\rm GC}}

\def\Pinsitu{P_{\rm in-situ}}

\def\Pin{P_{\rm in-situ}}
\def\Pacc{P_{\rm accreted}}
\def\Pthresh{P_{\rm thresh}}
\def\Nlayers{N_{\rm layers}}
\def\Nnodes{N_{\rm nodes}}
\def\Niter{N_{\rm iter}}

\def\apj{\rm Astrophys. J.}
\def\aap{\rm Astron. \&Astrophys.}
\def\mnras{\rm Mon.\,Not.\,RAS.}
\def\LCDM{$\Lambda$CDM }

\def\mathnew{\mathsurround=0pt}
\def\simov#1#2{\lower .5pt\vbox{\baselineskip0pt
    \lineskip-.5pt\ialign{$\mathnew#1\hfil##\hfil$\crcr#2\crcr\sim\crcr}}}

\def\apj{\rm ApJ}
\def\apjl{\rm ApJL}
\def\apjs{\rm ApJS}
\def\aj{\rm AJ}
\def\mnras{\rm MNRAS}
\def\nat{\rm Nature}

\def\aap{\rm A\&A}
\def\araa{\rm ARAA}

\defcitealias{Reina-Campos17}{RK17}
\defcitealias{Kruijssen12d}{K12}
\defcitealias{Trujillo-Gomez19}{TRK19}

\title[Inferring GC origins from observations]{In-situ or accreted? Using deep learning to infer the origin of extragalactic globular clusters from observables}

\author[S. Trujillo-Gomez et al.]
{Sebastian Trujillo-Gomez$^{1}$\thanks{E-mail: strujill@gmail.com}, 
J.~M.~Diederik Kruijssen$^{2}$, Joel Pfeffer$^{3}$, 
\newauthor
Marta Reina-Campos$^{4,5}$, Robert A. Crain$^{6}$, Nate Bastian$^{7,8}$ and Ivan Cabrera-Ziri$^{1}$
\\
$^{1}$Astronomisches Rechen-Institut, Zentrum f{\"u}r Astronomie der Universit{\"a}t Heidelberg, Monchhofstra{\ss}e 12-14, D-69120 Heidelberg, Germany \\
$^{2}$Cosmic Origins Of Life (COOL) Research DAO, coolresearch.io \\
$^{3}$International Centre for Radio Astronomy Research (ICRAR), M468, University of Western Australia, 35 Stirling Hwy, Crawley, WA 6009, Australia \\
$^{4}$ Department of Physics \& Astronomy, McMaster University, 1280 Main Street West, Hamilton, L8S 4M1, Canada\\
$^{5}$Canadian Institute for Theoretical Astrophysics (CITA), University of Toronto, 60 St George St, Toronto, M5S 3H8, Canada \\
$^{6}$Astrophysics Research Institute, Liverpool John Moores University, 146 Brownlow Hill, Liverpool L3 5RF, UK \\
$^{7}$Donostia International Physics Center (DIPC), Paseo Manuel de Lardizabal, 4, E-20018 Donostia-San Sebasti{\'a}n, Guipuzkoa, Spain \\
$^8$IKERBASQUE, Basque Foundation for Science, E-48013 Bilbao, Spain \\
}

\date{Accepted X{\sevensize xxxx} XX. Received 2023 January XX; in original form 2023 January 12}

\pubyear{2023}

\begin{document}
\label{firstpage}
\pagerange{\pageref{firstpage}--\pageref{lastpage}}
\maketitle

\begin{abstract}
Globular clusters (GCs) are powerful tracers of the galaxy assembly process, and have already been used to obtain a detailed picture of the progenitors of the Milky Way. Using the E-MOSAICS cosmological simulation of a $(34.4\Mpc)^3$ volume that follows the formation and co-evolution of galaxies and their star cluster populations, we develop a method to link the origin of GCs to their observable properties. We capture this complex link using a supervised deep learning algorithm trained on the simulations, and predict the origin of individual GCs (whether they formed in the main progenitor or were accreted from satellites) based solely on \emph{extragalactic} observables. An artificial neural network classifier trained on $\sim50,000$ GCs hosted by $\sim 700$ simulated galaxies successfully predicts the origin of GCs in the test set with a mean accuracy of $89$ per cent for the objects with $\feh<-0.5$ that have unambiguous classifications. The network relies mostly on the alpha-element abundances, metallicities, projected positions, and projected angular momenta of the clusters to predict their origin. A real-world test using the known progenitor associations of the Milky Way GCs achieves up to $90$ per cent accuracy, and successfully identifies as accreted most of the GCs in the inner Galaxy associated to the \emph{Kraken} progenitor, as well as all the \emph{Gaia-Enceladus} GCs. We demonstrate that the model is robust to observational uncertainties, and develop a method to predict the classification accuracy across observed galaxies. The classifier can be optimized for available observables (e.g. to improve the accuracy by including GC ages), making it a valuable tool to reconstruct the assembly histories of galaxies in upcoming wide-field surveys. 
\end{abstract}




\section{Introduction}
\label{sec:intro}

One of the major goals of astrophysics is understanding the physical processes that gave rise to galaxies out of the tiny density perturbations that emerged from the epoch of recombination. The advent of modern cosmology has provided precise knowledge of these initial conditions in the context of the highly successful $\Lambda$ Cold Dark Matter ($\Lambda$CDM) cosmological paradigm. The \LCDM model makes detailed predictions for the formation and evolution of dark matter (DM) haloes, which are the sites for baryonic material to condense into the galaxies we observe today \citep{Blumenthal84,Navarro95,Springel05}. Hydrodynamical cosmological simulations in the \LCDM framework predict that galaxies assemble through a combination of in-situ star formation in cold gas that is continuously accreted from cosmic filaments, and continuous infall of smaller satellite galaxies along with their DM, gas, and stars \citep{NaabOstriker17}. Sophisticated dynamical models of external galaxies using integral field spectroscopic data are now able to recover the properties of kinematically and chemically distinct `cold' and `hot' populations that trace the in-situ and accreted stellar components, respectively \citep[e.g.][]{Zhu20,Poci21}. However, reconstructing the detailed merger history of a galaxy from observations remains an extremely challenging task. 

The first chemo-dynamical studies of galaxies date back to the 1960s, when observations of the stars and globular cluster (GC) populations of the Milky Way (MW) showed that the kinematics of stars contained important clues to the origin of the various components. This pioneering work by \citet{Eggen62} found that while young stars follow nearly circular orbits, older stars have eccentric radial orbits with lower angular momentum and higher vertical velocity dispersion, all indicative of their accreted origin. Later studies of the ages and metal abundances of GCs in the Milky Way showed that while inner GCs follow a tight age-metallicity relation, the outer GCs have a broad range of ages at fixed metallicity \citep{SearleZinn78}. This simple observation confirmed the scenario where the MW disc stars and GCs formed early, while the halo formed slowly from material that continued to accrete long after the disc was in place. Several decades later, the Sloan Digital Sky Survey \citep[SDSS;][]{York00} found the first evidence of a past accretion event in the Milky Way, the Sagittarius stream, a remnant of the accretion of the \emph{Sagittarius} dwarf galaxy \citep{Ibata94}. 

The \emph{Gaia} survey \citep{GaiaDR2} revolutionized the field of Galactic archaeology by precisely measuring the 3D positions and motions of millions of stars in the inner halo, enabling the search for the progenitor galaxies of the Milky Way (MW) using the phase-space clustering of halo stars \citep[for a review, see][]{Helmi20}. In the last five years these data, combined with other spectroscopic surveys, led to the identification of the stellar debris from one of the most massive galaxy ever accreted by the Milky Way, \emph{Gaia-Enceladus} (also known as the \emph{Gaia Sausage}) \citep{enceladus,Haywood18,Belokurov18}, and of at least six additional progenitors \citep[e.g.,][]{Deason19, Myeong18a, Myeong18b, Myeong18d, sequoia, Iorio19, Koppelman19, thamnos, Mackereth19, Necib20a, Necib20b, Vasiliev19, Gallart19, Horta21, Malhan22}. Their location in 6D phase-space together with their metallicities and alpha-element abundances, identifies these substructures as having formed in satellites with different masses and star formation histories \citep[see][]{Helmi20}. The new data therefore allowed the global properties (such as mass and accretion redshift) to be determined for the most massive progenitors of the Galaxy. More recently, the H3 survey \citep{Conroy19} of high latitude stars in the MW found evidence of six chemo-dynamical substructures in the outer halo, beyond the reach of \emph{Gaia} \citep{Naidu20}. This brought the census of Galactic progenitors up to $\sim10$, accounting for $\sim 95$ per cent of the mass of the stellar halo. Achieving a similarly detailed assembly reconstruction for large samples of galaxies would undoubtedly open an entirely new window into galaxy formation and cosmology. 

\emph{Gaia} also provided the precise orbits of nearly all of the Galactic globular clusters \citep{Helmi18,Vasiliev19,Baumgardt19}. These data, along with the GC chemical abundances and ages, offered a novel and complementary way of reconstructing galaxy assembly. GCs are particularly powerful tracers of galaxy assembly because they can be studied at much larger distances than individual stars, up to $\sim 100\Mpc$, have long phase-mixing timescales, and their abundance relative to field stars increases in low-mass galaxies \citep{Peng08,Georgiev10,Forbes18b}. Using hydrodynamical cosmological simulations from the E-MOSAICS project, which include the formation and evolution of star clusters, \citet{emosaicsII} demonstrated that GCs are excellent tracers of the properties of their progenitor galaxies. \citet{krakenI} then used the age-metallicity relation of the MW GCs to obtain the most detailed reconstruction to date of the merger tree of the Galaxy. \citet{Trujillo-Gomez21a} found a surprising amount of galaxy assembly information encoded in the 3D GC system kinematics of simulated MW-mass galaxies, and applied a statistical method to the \emph{Gaia} data to produce an independent and consistent reconstruction of the MW merger tree. \citet{Massari19} used phase-space and age-metallicity information to associate most of the accreted GCs to each of the five most massive (likely) progenitors, \emph{Gaia-Enceladus}, \emph{Kraken}, \emph{Sagittarius}, \emph{Sequoia}, and the progenitor of the \emph{Helmi streams}. \citet{Pfeffer20} studied the relationship between the current phase-space distribution of GCs and the properties of their progenitors in cosmological simulations. Using machine learning to exploit this relation, along with the GC ages and metallicities in the simulations, \citet{krakenII} trained an artificial neural network to recover the masses and accretion redshifts of the five dominant MW progenitors. New studies continue to uncover further details of the MW assembly. For instance, \citet{Malhan22} analyzed the statistical 6D distribution of a large population of tracers in the Galactic halo (including GCs and stellar streams) to robustly search for phase-space substructures, and discovered a potential additional progenitor named \emph{Pontus}.

In this study, we aim to provide the initial steps to extend the powerful methods that have been applied to the Milky Way to recover the assembly histories of external galaxies based on their observed GC populations. First, we study the relation between the fraction of GCs accreted from satellites and fundamental galaxy properties in the simulations. We then investigate the link between GC origin (whether a GC was formed in-situ within the galaxy or was accreted), and its individual properties as determined by standard photometric and spectroscopic observations. The main result we highlight is that extragalactic GC observables contain a record of their progenitor properties, and that this information can be used to recover the origin of individual GCs using only a few key observables (their positions, radial velocities, and metallicities, and the stellar mass and effective radius of their host galaxy). With the goal of applying our classifier algorithm to upcoming deep, wide-field spectroscopic galaxy surveys, we provide the classification model code in a public repository. 

The paper is organized as follows. Section~\ref{sec:sample} describes the simulations and the galaxy and GC sample selection. Section~\ref{sec:properties} shows how the accreted fraction depends on galaxy properties. Section~\ref{sec:model} describes the deep learning model used to predict GC origin in external galaxies, and Section~\ref{sec:results} provides a detailed analysis of its predictions for simulated galaxies, and the results of the a real-world test using the MW data. The results and discussed in Section~\ref{sec:discussion}, and summarized in Section~\ref{sec:conclusions}.

\section{Simulated galaxy and GC sample}
\label{sec:sample}

In this work we use the simulated galaxies and star cluster populations from the E-MOSAICS simulations. Below we describe the simulations and sample selection criteria. 

\subsection{The E-MOSAICS simulations}
\label{sec:simulations}

E-MOSAICS (MOdelling Star cluster population Assembly In Cosmological Simulations within EAGLE) is a suite of hydrodynamical cosmological simulations that follow the formation and co-evolution of galaxies and their star cluster populations \citep{emosaicsI,emosaicsII}. The physics of galaxy formation is implemented using the EAGLE model \citep{Schaye15,Crain15}, which uses a feedback prescription calibrated to reproduce the stellar mass function and disc-galaxy sizes at $z=0$. The EAGLE model also reproduces many additional key properties of the observed galaxy population, including their present-day luminosities and colours \citep{Trayford15}, the evolution of the stellar mass function, star formation rates \citep{Furlong15}, and galaxy sizes \citep{Furlong17}, and the chemical abundances of stars in the Milky Way \citep{Mackereth18}. 

To model the formation and evolution of star clusters, the simulations use an improved version of the MOSAICS subgrid model \citep{Kruijssen11,emosaicsI}. Star clusters are treated as a subgrid population within each star particle, and form according to an environmentally-dependent prescription based on models for the fraction of stars formed in bound clusters \citep{Kruijssen12b}, and for the upper truncation mass of the Schechter initial cluster mass function \citep[ICMF;][]{Reina-Campos17}. Both of these quantities are calculated using the local gas conditions, and increase with the gas pressure. Clusters lose mass via stellar evolution, two-body relaxation, and tidal shocks, and may be completely disrupted by infall into the centres of galaxies via dynamical friction. Mass loss due to tidal shocks and two-body relaxation is calculated self-consistently at each time step from the local tidal field. 

The E-MOSAICS simulations have been shown to reproduce several key properties of GC populations. These include the massive end of the GC mass function \citep{emosaicsI,Hughes22}, GC specific frequencies \citep{emosaicsII,Bastian20}, the color-luminosity relation of metal-poor GCs \citep[the `blue tilt',][]{Usher18}, the  GC radial distribution \citep{Reina-Campos21} and kinematics \citep{Trujillo-Gomez21a}, and the GC system mass-halo mass relation \citep{Bastian20}. They also reproduce the age distribution of GCs in satellite streams \citep{Hughes19a}, and the fraction of stars in the bulge of the Galaxy that were born in GCs \citep{Hughes19b}. These simulations demonstrated that the properties of GC populations reflect the environment and assembly of their host galaxies \citep{emosaicsII}, and this allowed the most detailed reconstruction so far of the merger tree of the Milky Way \citep{krakenI}, including the prediction of the masses and accretion times of its five most massive progenitors using the properties of its GCs \citep{krakenII}. We refer the reader to \citet{emosaicsI} for a complete description of the physical models in the simulations.

The E-MOSAICS simulations are unique in their ability to model star cluster populations in a cosmological volume to $z=0$, and their success in reproducing galaxy and GC observables makes them an ideal tool to investigate how the intrinsic properties of GCs relate to their natal galaxies. In this work we use the galaxies and GCs from the E-MOSAICS (34.4 cMpc)$^3$ periodic volume \citep{Bastian20}. The gas particle mass is $2.26\times10^5\Msun$, and the gravitational softening at $z=0$ is $\epsilon = 0.35\kpc$. An FoF algorithm \citep[Friends-of-Friends,][]{Davis85} is first used to identify DM groups with a linking length of 0.2 times the mean particle separation. Within each group, the {\sc SubFind} algorithm \citep{springeletal01,Dolag09} then identifies gravitationally bound structures, and identifies as the central subhalo/galaxy the one containing the particle with the lowest potential energy. All other subhaloes within the FoF group are then considered satellites of the central galaxy. Merger trees are constructed in the same way as for the EAGLE simulations, using the {\sc D-Trees} algorithm \citep{Jiang14,Qu17} to link between 10 and 100 of the most bound particles in each subhalo across the 28 snapshots \citep[for further details on this procedure, see][]{Qu17}. The simulation volume contains 2900 galaxies (resolved with at least 100 star particles), and [465, 69, 7] galaxies with $\Mhalo > [10^{11} , 10^{12} , 10^{13}]\Msun$. We refer the reader to \citet{Bastian20} for a detailed description of the simulation and its first results.

\subsection{Sample selection}
\label{sec:sample_selection}

For the analysis in this study we select all central galaxies (i.e. not satellites) in the (34.4 cMpc)$^3$ periodic volume with stellar masses $\Mstar>10^8\Msun$ and hosting at least 10 GCs. GCs are identified as star clusters with $M>M_{\rm thresh}$ and metallicities $-2.5 < \rm{[Fe/H]} < \rm{[Fe/H]}_{\rm thresh}$. The minimum mass threshold is chosen to increase with galaxy mass to follow the shift in the upper truncation of the GC mass function \citep{Hughes22}. The metallicity range is designed to mitigate the effect of numerical cluster underdisruption due to the absence of cold and dense gas in the EAGLE model \citep{Reina-Campos21}. It effectively removes most of the excess GCs (which are mostly metal-rich) that should have been effectively disrupted by tidal shocks \citep[for a detailed discussion see Appendix D of][]{emosaicsII}. To emulate the observationally motivated galaxy mass-dependent GC minimum mass used by \citet{Hughes22}, we use the smoothly varying function 
\begin{equation}
    \log(M_{\rm thresh}) = 0.5\log(\Mstar) - 0.5 .
\end{equation}
As a result, for the lowest mass galaxies in the sample we select as GCs star clusters with $M > 10^{3.5}\Msun$, while for the most massive ellipticals we use $M \ga 10^{5}\Msun$. Table~\ref{tab:table1} shows the upper metallicity threshold values as a function of galaxy stellar mass.

\begin{table}
\centering{
  \caption{Upper metallicity thresholds used to define the GC sample as a function of host galaxy stellar mass. The upper threshold is designed to remove artificially underdisrupted clusters (see Sect.~\ref{sec:sample_selection}). The last column shows the metallicity used to split the GC sample into metal-poor and metal-rich subpopulations.}
  \label{tab:table1}
	\begin{tabular}{ccc}\hline\hline
		galaxy $\log(\Mstar/\Msun)$ & [Fe/H]$_{\rm thresh}$ & [Fe/H]$_{\rm split}$  \\ \hline
		8.0 - 8.5   & -1.0 & -1.2 \\
		8.5 - 9.0   & -1.1 & -1.2 \\
		9.0 - 9.5   & -0.8 & -1.2 \\
		9.5 - 10.0  & -0.5 & -1.1 \\
		10.0 - 10.5 & -0.5 & -1.0 \\
		10.5 - 11.0 & -0.5 & -0.9 \\
		$>11.0$     & -0.3 & -0.8 \\
		\hline \hline
	\end{tabular}}
\end{table}

These criteria result in a sample of 921 central galaxies hosting a total of 75,810 GCs. To classify the GCs into in-situ and accreted, we use the merger trees and examine the two snapshots that bracket the formation time of the host star particle. If the progenitor gas particle was assigned to the same branch of the merger tree as the resulting star particle, the GC has a clear origin, and it is labelled `in-situ' if it formed on the main branch, or `accreted' (or `ex-situ') if it formed in a different branch. For those GCs with a formation time that falls between two different branches, the origin is unclear (given the spacing between simulation snapshots), and they are labelled `ambiguous'. After classification, the sample contains 39,158 in-situ, 36,652 accreted, and 6,674 ambiguous GCs. Since ambiguous GCs are simply an artefact of the merger trees, we remove them from the final sample. The final sample contains 57 per cent in-situ, and 43 per cent accreted GCs.

\section{Globular cluster origin across the galaxy population}
\label{sec:properties}

\begin{figure}
    \includegraphics[width=\columnwidth]{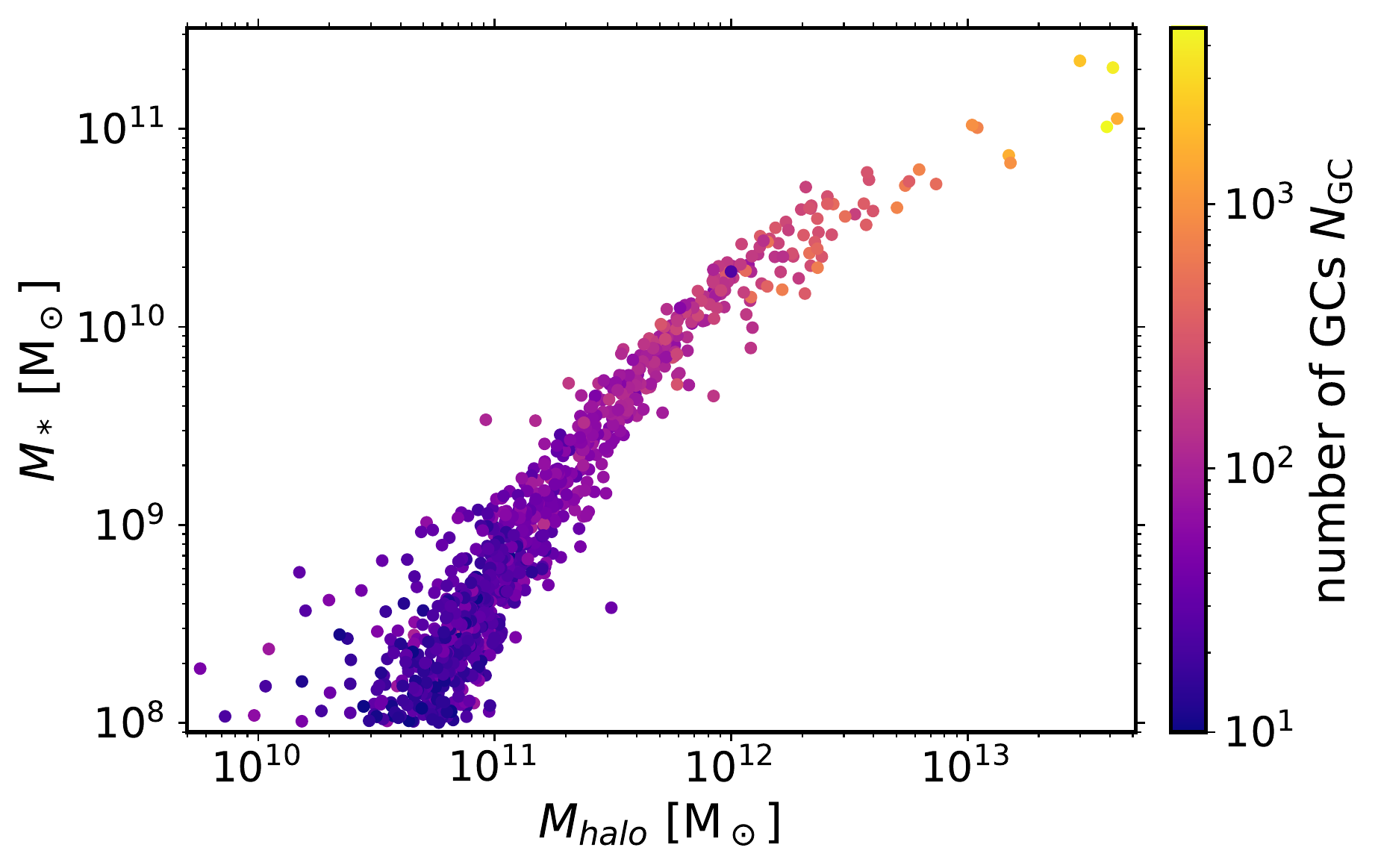}
    \caption{Number of GCs hosted by each simulated galaxy as a function of the galaxy stellar and halo mass. The number of hosted GCs increases steeply with both stellar and halo mass. At fixed halo mass, galaxies with larger stellar mass tend to host more GCs. See Section~\ref{sec:sample_selection} for the sample selection criteria.}
\label{fig:SMHM}
\end{figure}

We begin by examining how GC origin is related to GC and host galaxy properties. Figure~\ref{fig:SMHM} shows the stellar-to-halo mass relation of the simulated galaxies coloured by the number of GCs they host. The size of the GC population increases steeply with halo mass, reproducing the observed qualitative trend \citep[e.g.][]{Blakeslee97,Burkert20}. In a more detailed analysis, we found that the relation between halo mass and total mass in GCs in the simulations also matches observations \citep{Bastian20}. At fixed galaxy stellar mass there is a weak secondary trend of increasing number of GCs with increasing halo mass.

\begin{figure}
    \includegraphics[width=\columnwidth]{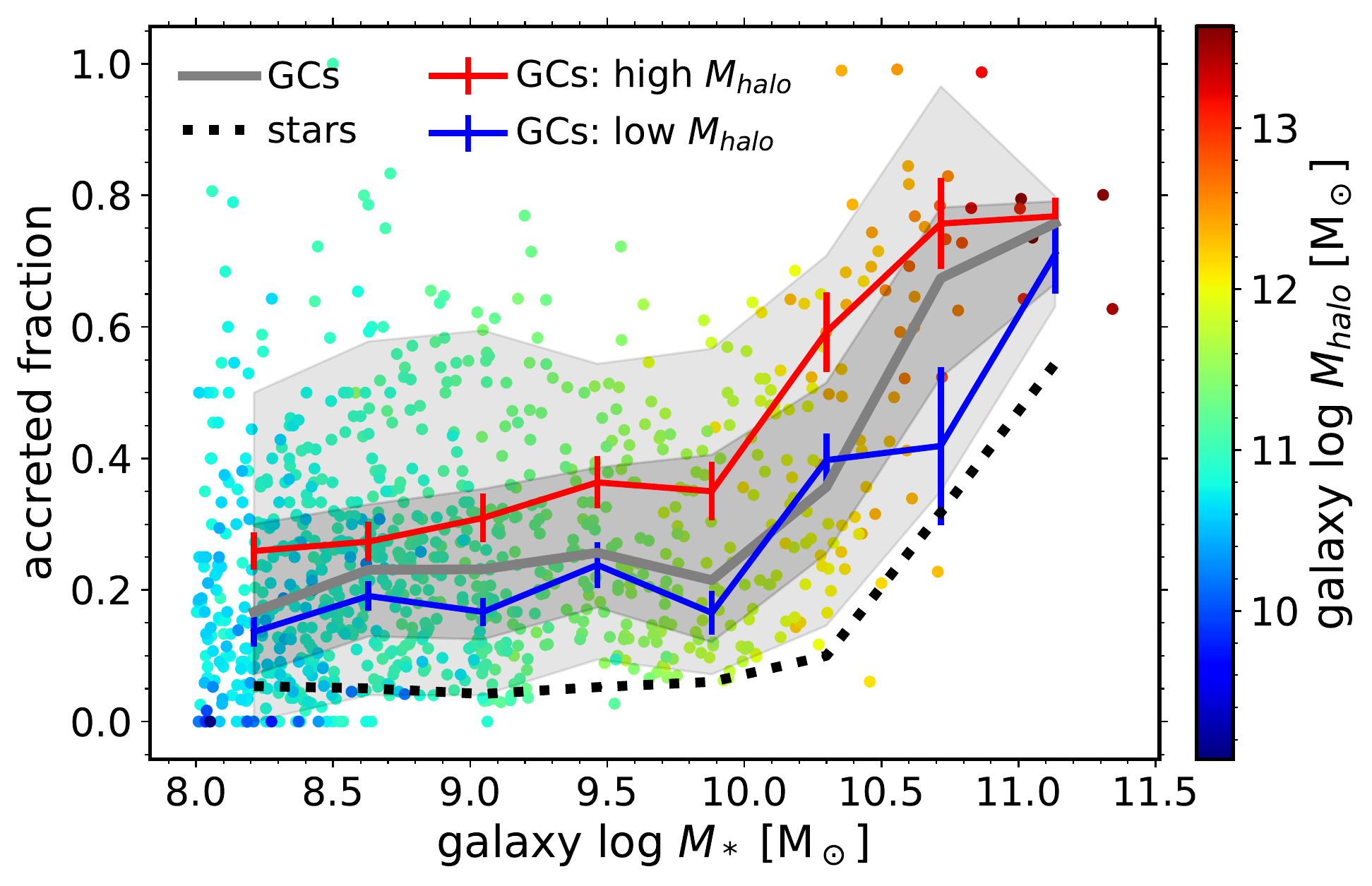}
    \caption{Fraction of stars and GCs accreted from satellites as a function of host galaxy stellar mass (points), coloured by halo mass. The grey line and shading show the median, [5,95], and [25,75] percentile range of the accreted fraction in bins of stellar mass (with error bars corresponding to the uncertainty in the median). The blue and red lines show the median accreted fraction in the bottom and top quartiles of halo mass in each bin, respectively. The dotted line shows the fraction of accreted stars. The median GC accreted fraction increases with stellar mass, and is always larger than for stars. At fixed stellar mass, galaxies hosted by more massive DM haloes have larger fractions of accreted GCs.  }
\label{fig:accretedfrac_Mhalo}
\end{figure}

Figure~\ref{fig:accretedfrac_Mhalo} shows the fraction of accreted GCs and stars in each galaxy as a function of galaxy stellar mass. The fraction of accreted GCs in E-MOSAICS increases with galaxy stellar mass following the qualitative trend found for stars in semi-empirical and semi-analytical models, as well as in cosmological simulations including EAGLE \citep[e.g.][]{Rodriguez-Gomez16,Qu17,Clauwens18,Tacchella19,Davison20,Moster20}. This is not surprising, and is a direct consequence of the hierarchical nature of galaxy assembly combined with the shape of the fundamental stellar-to-halo mass relation (Fig.~\ref{fig:SMHM}). The stellar-to-halo mass relation is very steep at low masses and becomes shallower for galaxies more massive than the MW. Massive galaxies are therefore partially assembled by hierarchical accretion of satellites with relatively high stellar masses, while dwarfs accrete only satellites with relatively low stellar masses. While the accreted fraction increases with galaxy mass for both stars and GCs, Fig.~\ref{fig:accretedfrac_Mhalo} shows that the fraction of accreted GCs is always larger. This is a result of the higher mean specific frequencies of satellites relative to centrals.

\begin{figure*}
	\includegraphics[width=\columnwidth]{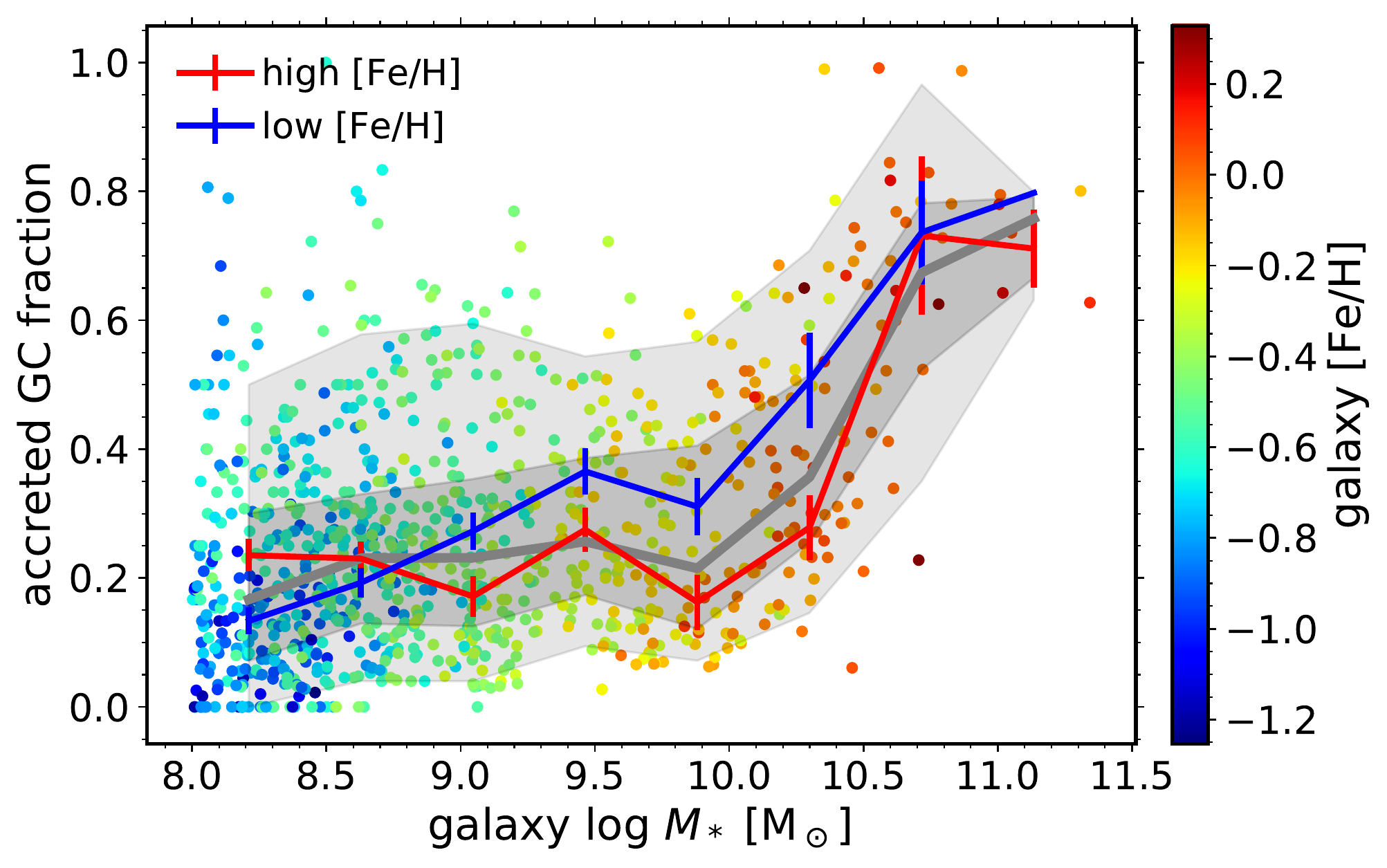}
    \includegraphics[width=\columnwidth]{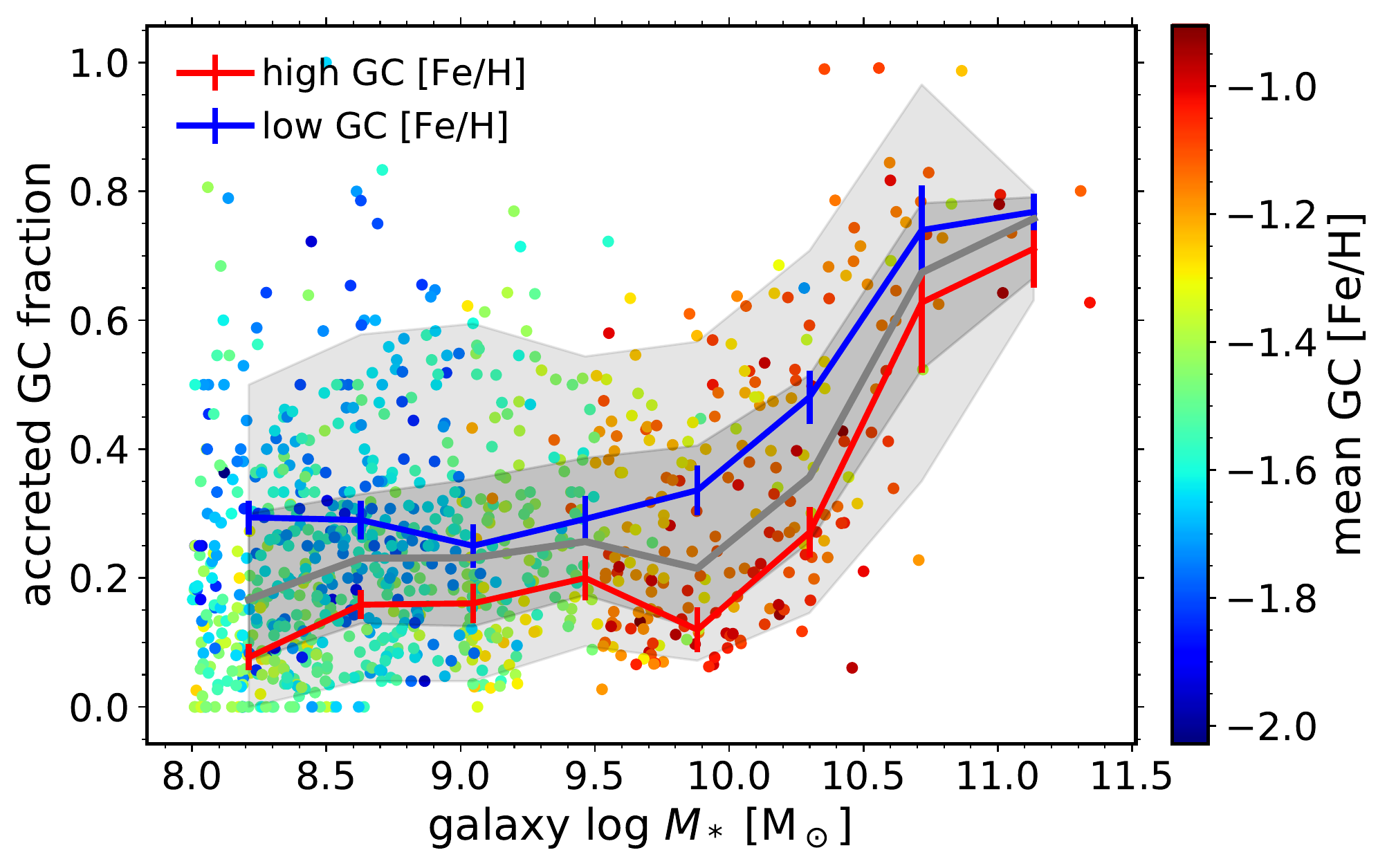}
    \caption{Accreted GC fraction as a function of galaxy stellar mass. Left: coloured by mean galaxy metallicity [Fe/H]. The blue and red lines show the median accreted GC fractions for the galaxies with the 25 per cent lowest and highest metallicities, respectively. Right: coloured by mean GC metallicity. The blue and red lines show the median accreted GC fractions for the galaxies in the bottom and top GC system metallicity quartiles, respectively. At fixed stellar mass, galaxies with higher metallicity stars and GCs have systematically lower accreted fractions than those at lower metallicities. This is driven by the fact that accreted stars and GCs from low-mass satellites are metal-poor.}
\label{fig:accretedfrac_feh}
\end{figure*}

There is significant scatter in the GC accreted fraction at fixed galaxy stellar mass. To understand the physical drivers of the scatter, we search for secondary trends in the GC accreted fraction. Figure~\ref{fig:accretedfrac_Mhalo} also shows the median GC accreted fraction of galaxies hosted by the least/most massive DM haloes in each stellar mass bin (in the lower/upper quartile of the distribution of $\Mhalo$ in each bin). At fixed stellar mass, galaxies hosted by more massive DM haloes have larger GC accreted fractions, as expected from their larger fraction of accreted material from DM-rich satellites. The left panel of Figure~\ref{fig:accretedfrac_feh} shows how the accreted fraction of stars varies with galaxy metallicity, with metal-poor galaxies typically hosting a larger fraction of accreted stars. This trend is driven by the decrease in the mean metallicity due to the accretion of a larger fraction of stars from satellites\footnote{There is an additional weak trend where the in-situ GCs in massive galaxies with high GC accreted fractions tend to be less enriched than in those with low accreted fractions. This originates from the anticorrelation between halo mass (or formation time) and galaxy metallicity at fixed stellar mass discussed above.}. We also find that the mean metallicity of accreted GCs is higher in galaxies with higher GC accreted fractions due to the dominant contribution of the most massive satellite (which also contains the most metal-rich GCs). 

In the right panel of Figure~\ref{fig:accretedfrac_feh} we show an even stronger trend found in the accreted fraction of metal-poor and metal-rich GC systems (i.e. considering the mean GC metallicity of each galaxy). At fixed stellar mass, galaxies with metal-poor GC systems have systematically higher accreted GC fractions compared to those with metal-rich systems. As in the case of the stellar component, this distinct metallicity dependence of the accreted fraction results from a combination of the overall effect of larger fractions of (metal-poor) GCs accreted from satellites on the mean GC metallicity, and the effect of DM halo formation times on the in-situ GCs.

\begin{figure}
    \includegraphics[width=\columnwidth]{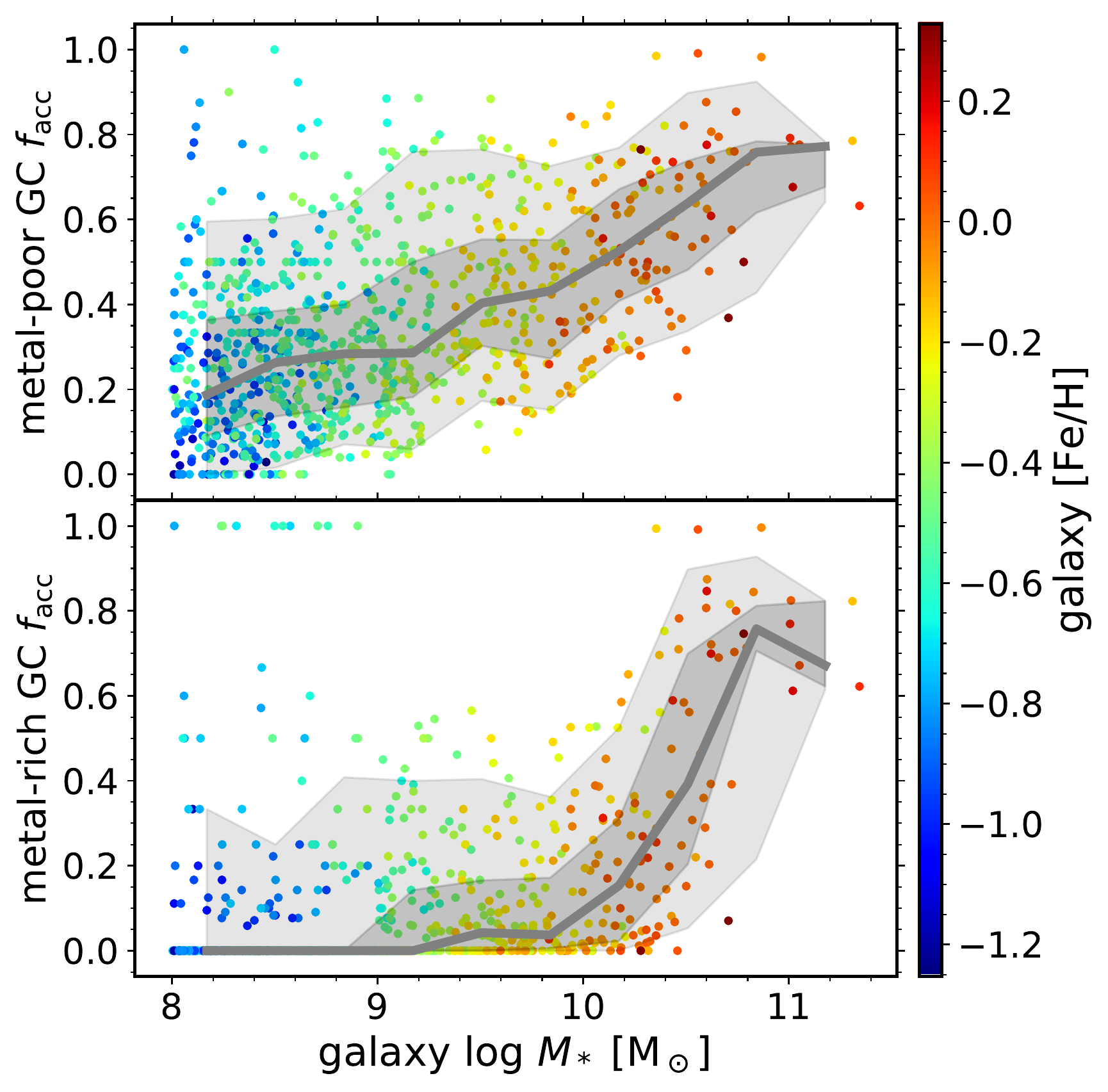}
    \caption{Accreted GC fraction as a function of galaxy stellar mass for metal-poor (top panel) and metal-rich (bottom panel) GC populations. The GC subpopulations are selected based on the stellar mass-dependent metallicity split in Table~\ref{tab:table1}. Metal-poor GCs typically have a mixed origin. Metal-rich GCs in low-mass galaxies are almost exclusively formed in-situ, while in galaxies more massive than the MW they have a mixed origin.}
\label{fig:accretedfrac_GCmetbins}
\end{figure}

\begin{figure}
    \includegraphics[width=\columnwidth]{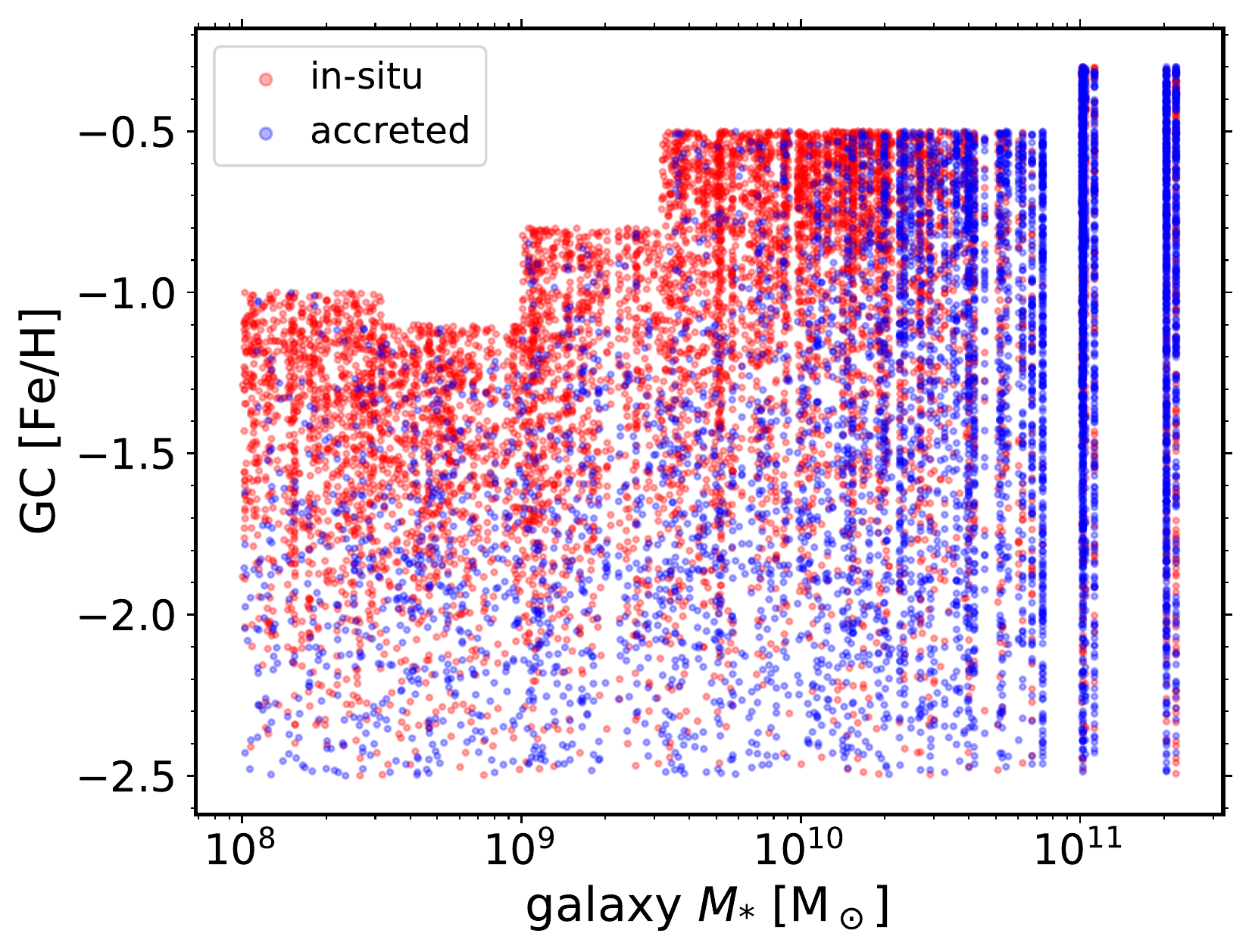}
    \caption{Origin of individual GCs in the simulation as a function of GC metallicity and host galaxy mass. The upper mass-dependent metallicity limit reflects the selection applied to reduce contamination by artificially underdisrupted GCs (see Section~\ref{sec:sample_selection}). Accreted GCs tend to have lower metallicities than in-situ GCs, but there is a significant overlap between the two populations. There is also significant variation in the metallicity distribution of the accreted and in-situ GCs across galaxies of similar mass due to differences in their assembly histories.}
\label{fig:GC_origin_vs_feh}
\end{figure}

Figure~\ref{fig:accretedfrac_GCmetbins} shows the accreted fraction of metal-poor and metal-rich GC subpopulations as a function of galaxy stellar mass. The subpopulations are defined using the stellar mass-dependent split shown in Table~\ref{tab:table1}. Metallicity alone is not a direct proxy for GC origin. While metal-rich GCs tend to form in-situ in low-mass galaxies, the metal-poor population is typically a mix of in-situ and accreted objects. Figure~\ref{fig:GC_origin_vs_feh} shows the distribution of accreted and in-situ GCs as a function of GC metallicity and host galaxy mass. It confirms that while accreted GCs tend to be more metal-poor than in-situ GCs, there is significant overlap and galaxy-to-galaxy variation in the populations, and metallicity alone is generally not enough to determine GC origin.

\section{Predicting GC origin using Machine Learning}
\label{sec:model}

We now turn to the question of whether the origin of a particular GC can be predicted using its observable properties, and which observables are best suited for this purpose. We take advantage of the flexibility and predictive power of deep learning algorithms when applied to problems with highly nonlinear relations between the input and output variables. In addition, we explore other supervised learning techniques to find possible alternatives with higher predictive power. 

After exploring several classifier algorithms including k-nearest neighbors \citep{FixHodges89}, Logistic Regression \citep{PearlReed1920}, Support Vector Machines \citep{Boser92}, Decision Trees \citep{Hunt66}, and Random Forests \citep{Breiman01}, we find that their predictive accuracy is generally lower compared to deep learning, while most do not provide probabilistic outputs. The probabilistic output of neural networks will be key for tuning and predicting the uncertainties in the model (see Sections \ref{sec:performance} and \ref{sec:confidence}).

\subsection{Algorithm description}
\label{sec:algorithm_description}

For the fiducial model we employ a Multilayer Perceptron artificial neural network\footnote{also known as feed-forward neural network.} architecture \citep[MLP;][]{Rumelhart86} with dense, sequential layers. MLPs are powerful classifiers that are ideally suited for complex problems where the classes are not linearly separable, as we expect here. They are also advantageous compared to more traditional models because they automatically create useful new features from the provided inputs. We use the deep learning library {\sc Keras} \citep{keras} implemented within the {\sc TensorFlow} framework \citep{tensorflow}. 

The MLP architecture consists of several layers of artificial neurons that are connected in sequence, such that each neuron takes as input the combined outputs from all the neurons in the previous layer. To adapt the model to our specific classification task, we set the input layer to contain as many nodes (i.e. neurons) as the dimensions of the input data (i.e. the number of GC observables, $N_{\rm input}$), and the output layer to have 2 dimensions corresponding to the two possible classification labels: \emph{in-situ} and \emph{accreted}. The number of hidden layers $\Nlayers$, and the number of nodes per layer ($\Nnodes$) are left as free parameters to be optimized using the validation data. The input and hidden layers use the standard `Rectified Linear Unit' (ReLU) activation function, $h(x)=\max(0,x)$, and the output layer uses the sigmoid activation function to convert the output into a binary probability in the range $[0,1]$, $\Pin = 1-\Pacc$. The model is compiled using the `Adam' optimizer \citep{Kingma14}, with the standard binary cross-entropy loss function used in binary classification tasks,
\begin{equation}
  \mathcal{L} = -\frac{1}{\Ngc} \sum_{i=1}^{\Ngc} y_i\log(P_i) + (1-y_i)\log(1-P_i) ,
\end{equation}
where $y_i$ is the true label and $P_i$ is the output probability for GC $i$ ($1$ for in-situ, $0$ for accreted). Below we describe the input features and training procedure.

\subsection{Training a neural network classifier on simulated galaxies and their GCs}
\label{sec:training}

To select the set of input observables (i.e. the features) used by the model to predict GC origin, we first explore a large set of physically-motivated GC and host galaxy observables. We iteratively remove features that do not affect the accuracy of the predictions to reduce as much as possible the complexity of the model. This procedure yields a fiducial set of $N_{\rm input}=17$ observables that we use in the final step to optimize the artificial network (ANN) architecture and to train the fiducial model. Table~\ref{tab:table2} summarizes the features. These are all derived using physically-motivated combinations of GC observables (metallicity, alpha-element abundance, projected position on the sky, and line-of-sight velocity), and global galaxy properties (stellar mass, mean metallicity, effective radius, and stellar velocity dispersion). Since GC ages are notoriously difficult to measure precisely beyond the Milky Way, we ignore them here and evaluate their contribution to the predictions in Section~\ref{sec:ages}. 

We select a single random orientation for each galaxy corresponding to a projection onto the $x-y$ plane of the simulation box, and calculate the positions and velocities in the reference frame of the centre of the galaxy obtained using {\sc SubFind}. We define the GC `rotation velocity' as the dot product of the GC LOS velocity $\vb{\Vp}$ and the unit vector pointing in the direction of net rotation of the galaxy at the projected GC position, $\Vrot \equiv \vb{\Vp} \vdot \vb{V}_{\rm rot}^{\rm gal}/|\vb{V}_{\rm rot}^{\rm gal}|$. We further test an augmented feature set by including an additional set of six features that describe the distribution of the projected distance and LOS velocity of the GC system (using the median, inter-quartile range, skewness, and kurtosis), and quantify the projected GC distance and velocity relative to the four nearest GC neighbours. We find that these additional features do not increase the model performance, and therefore keep only the original set of 17 input observables. Having chosen the final feature set, we follow the common practice of standardizing each feature by subtracting the mean and diving by the standard deviation to obtain distributions with a mean of 0 and standard deviation of 1.  

\begin{table*}
\centering{
  \caption{GC and host galaxy observables used as features in the fiducial neural network classifier. Projected positions and LOS velocities are calculated with respect to the position and velocity of the centre of the galaxy, assuming a single random orientation for each galaxy.}
  \label{tab:table2}
	\begin{tabular}{lcl}\hline\hline
		Feature & Object & Definition  \\ 
		\hline
		$\log\Mstargal$ & galaxy & stellar mass \\
		$\log\Regal$ & galaxy & projected effective radius \\
		$\fehgal$ & galaxy & mean metallicity \\
		$\alphagal$ & galaxy & mean oxygen abundance relative to iron [O/Fe] \\
		$\sigmagal$ & galaxy & stellar velocity dispersion \\
		$\log\Ngc$ & galaxy & total number of GCs \\
		$\sigmagcs$ & galaxy & GC system velocity dispersion \\
		$\fehgc$ & GC & metallicity \\
		$\alphagc$ & GC & oxygen abundance relative to iron [O/Fe] \\
		$\deltafeh$ & GC/galaxy & metallicity relative to the galaxy, $\fehgc - \fehgal$ \\
		$\deltaalpha$ & GC/galaxy & alpha-abundance relative to the galaxy, $\alphagc - \alphagal$ \\
		$\log\Rp/\Regal$ & GC/galaxy &  projected distance from galaxy centre in units of the galaxy effective radius \\
		$\sqrt{|\Vp|/\sigmagal}$ & GC/galaxy & LOS velocity in units of the galaxy velocity dispersion \\
		$\sqrt{|\Vp|/\sigmagcs}$ & GC/galaxy & LOS velocity in units of the GC system velocity dispersion \\
		$\Vrot/\sigmagal$ & GC/galaxy & `projected rotation velocity': dot product of LOS velocity and the unit vector pointing along the galaxy rotation velocity at the GC projected position in units of the galaxy velocity dispersion (see Section~\ref{sec:training}) \\
		$\log\Rp|\Vp|$ & GC & `projected angular momentum': product of the projected galactocentric distance and magnitude of LOS velocity \\
		$(\Rp\Vrot)^{1/3}$ & GC/galaxy & `projected angular momentum vector': product of projected galactocentric distance and $\Vrot$ (see Section~\ref{sec:training}) \\
		\hline \hline
	\end{tabular}}
\end{table*}

To train the model we use the sample of 69,136 simulated GCs with unambiguous origin hosted by 921 central galaxies (see Section~\ref{sec:properties}). Normally, the model would be trained on a random subsample containing the majority of the simulated GCs (typically $\sim 70{-}80$ per cent), and the remaining fraction would be used in model validation and testing. However, to avoid leakage of the information on host galaxy properties from the GCs in the training set to the GCs in the test set, we adopt a different approach. Instead, we split the host galaxies randomly into a training set containing all the GCs hosted by a subset comprised of 80 per cent of the galaxies (50,612 GCs from 736 galaxies), and a test set containing all the GCs in the remaining 20 per cent (18,524 GCs from 185 galaxies). This ensures that the model is not exposed to any of the test data during training, and increases its capacity to generalize to other datasets, including GCs in the real Universe. 

After selecting the training and test sets, we perform a grid search to optimize the main hyperparameters of the network: the number of layers, and the number of nodes (neurons) per layer. To maximize the use of the simulation data, we choose to use the same data for both validation and testing. We have verified that using separate validation and test data has no effect on the ability of the model to generalize to new data.\footnote{We tested this by running an experiment where the model performance was tested using data that had not been used in the hyperparameter tuning. The accuracy was unaffected.} We evaluate the validation accuracy of predictions using the test set for a model with parameters in the two-dimensional grid defined by the values $\Nlayers \in [2,3,4,5,6,7,8,9,10]$, and $\Nnodes \in [10,20,50,100,200]$. In each iteration, the training is stopped after 30 epochs, or when the accuracy does not increase over 5 epochs. The architecture with $[\Nlayers,\Nnodes] = [4,20]$ results in the highest validation accuracy $\approx 80$ per cent, and we select it for the fiducial model. Using these parameters we retrain the final model for 100 epochs, stopping early when the accuracy does not increase over the last 20 epochs. This model is saved and used to evaluate the predictions and performance in Section~\ref{sec:results}. We refer to it throughout the paper as the `fiducial' model.

\section{Results}
\label{sec:results}

In this section we evaluate and tune the performance of the fiducial model on the simulated test data. We then analyze the detailed predictions and the relative importance of each observable, and evaluate the model confidence. We also perform the first real-world test of the algorithm by predicting the origin of the MW GCs. Lastly, we test the impact of observational uncertainties and evaluate the performance improvement when GC ages are included.

\subsection{Model performance}
\label{sec:performance}

\begin{figure}
    \includegraphics[width=\columnwidth]{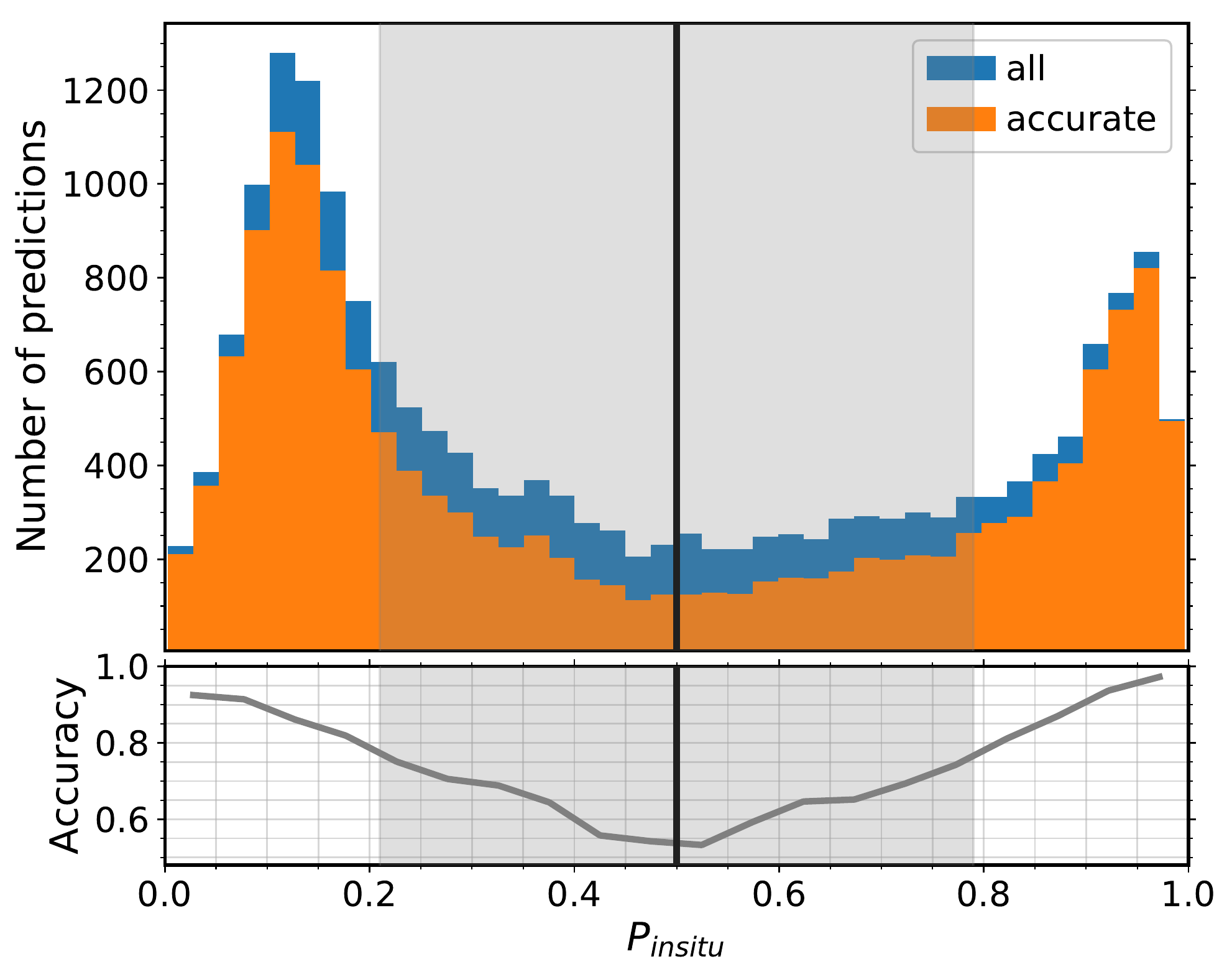}
    \caption{Accuracy distribution of the classifier on the simulated test sample as a function of the predicted probabilities. Top: distribution of the predicted in-situ probability $\Pin$ across the entire GC test sample compared to the distribution of only the correct predictions (assuming a decision threshold $\Pthresh=0.5$). Bottom: accuracy of the predictions in each probability bin. The vertical line marks the initial value of $\Pthresh=0.5$ we adopt for labelling the predictions of the classifier, where $\Pin > \Pthresh$ corresponds to in-situ, and $\Pin \leq \Pthresh$ corresponds to accreted. The grey shaded region indicates ambiguous predictions as defined in Sect. \ref{sec:performance}. Both the probability distribution and the accurate predictions are peaked near the two extremes of $\Pin$, showing that the classifier makes accurate predictions with high confidence.}
\label{fig:accuracy_Pinsitu}
\end{figure}

We now evaluate the performance of the classifier on the test sample containing 185 galaxies (i.e. 20 per cent of the sample) drawn at random from the simulation, and the 18,524 GCs they host (with unambiguous origin). Figure~\ref{fig:accuracy_Pinsitu} shows the distribution of predicted probabilities $\Pin$ for the simulated test sample (top panel) along with the number of correct predictions. To calculate the accuracy (i.e. the fraction of correct predictions), we must map the predicted probabilities output by the classifier to binary class labels assuming a simple probability threshold $\Pthresh = 0.5$, such that a GC is labelled `in-situ' when $\Pin>\Pthresh$, and `accreted' otherwise. The overall accuracy of the model (measured across the entire test GC sample) is $80$ per cent. We find that the classifier produces two distinct peaks in the distribution of probabilities, with each peak near the maximum probability for each class (i.e. $\Pin\sim0$ or $\Pin\sim1$). This shows that the neural network reaches a high confidence when predicting the origin of the majority of the GCs in the test sample. The fraction of correct predictions in each bin is shown in the bottom panel of Figure~\ref{fig:accuracy_Pinsitu}. The accuracy increases monotonically towards the most confident predictions (at $\Pin\sim0$, and $\Pin\sim1$). This is evidence that the classifier successfully predicts the correct labels with high confidence (i.e. high probabilities). A minority of the predictions lie in the ambiguous region with $\Pin \sim 0.3-0.7$.  

\begin{figure}
    \includegraphics[width=1.0\columnwidth]{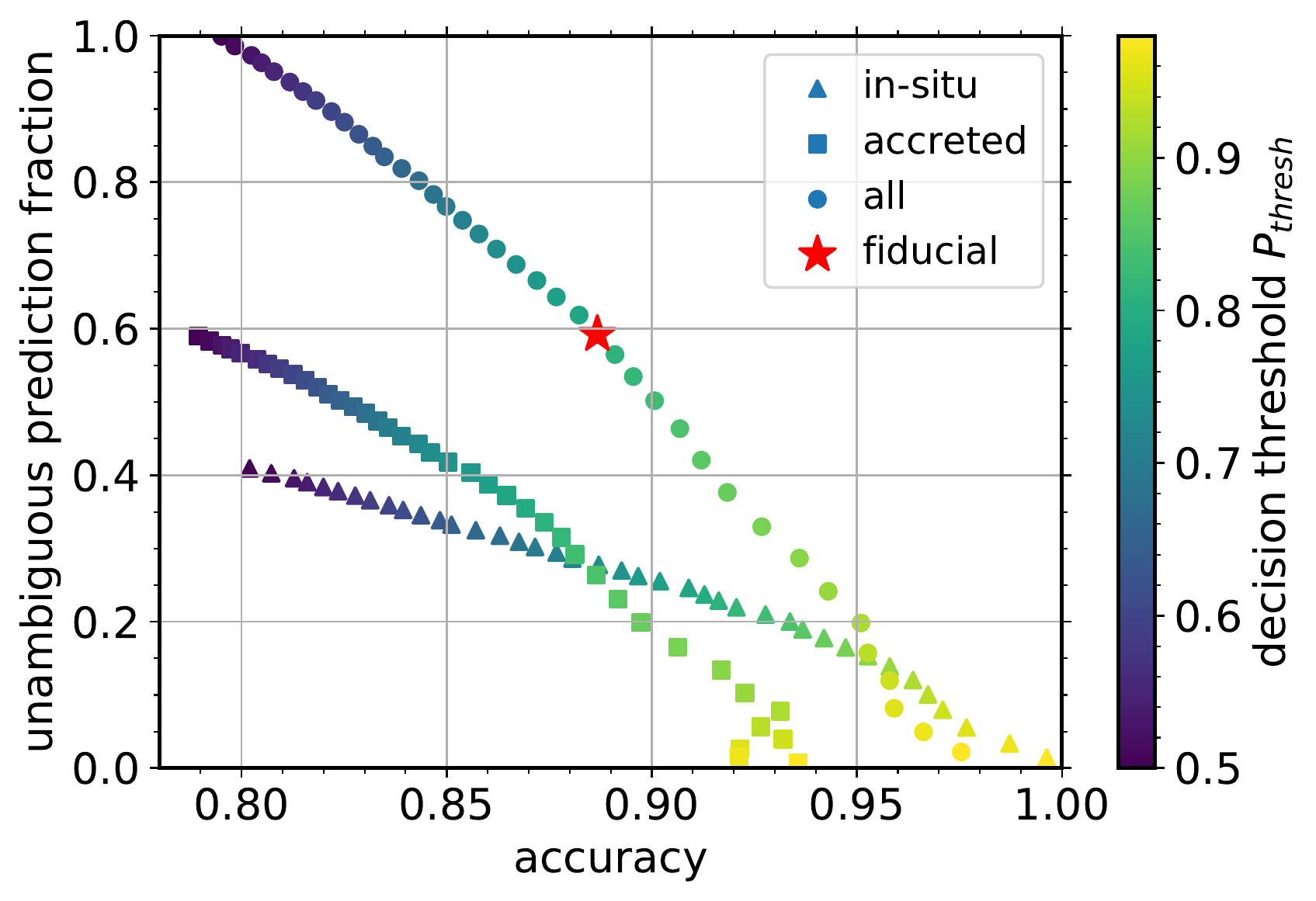}
    \caption{Effect of increasing the decision threshold of the neural network classifier on the fraction of unambiguous predictions in the test sample and their accuracy. The colour bar shows the decision threshold $\Pthresh$ for each class as well as for the combined sample. A GC is labelled `in-situ' when $\Pin > \Pthresh$, and `accreted' when $\Pacc \equiv 1-\Pin > \Pthresh$. The star symbol indicates the fiducial decision threshold adopted in this work. It corresponds to an accuracy of $\sim 89$ per cent on $60$ per cent of the GC sample.}
\label{fig:accuracy_samplefrac}
\end{figure}

To exploit the probabilistic nature of the model to improve the accuracy of the classifications, we introduce a new label, and define `ambiguous' predictions as those with $\Pin>\Pthresh$ and $\Pacc \equiv 1-\Pin>\Pthresh$, where $\Pthresh$ is the decision threshold. Figure~\ref{fig:accuracy_samplefrac} shows the effect of increasing the decision threshold on the fraction of unambiguous predictions, and on their accuracy. As expected, the accuracy increases with $\Pthresh$, while the completeness (i.e. the unambiguous fraction of predictions) decreases: $3/4$ of the sample reaches an accuracy of 85 per cent, while only half of the sample reaches 90 per cent accuracy. To optimize both the accuracy and the sample completeness, we define the unambiguous predictions using a fiducial value of $\Pthresh = 0.79$. This results in an accuracy of $\sim 89$ per cent (for a 60 per cent completeness). The ambiguous region is shown using grey shading in Figure~\ref{fig:accuracy_Pinsitu}.   

\begin{figure}
    \centering
    \includegraphics[width=0.9\columnwidth]{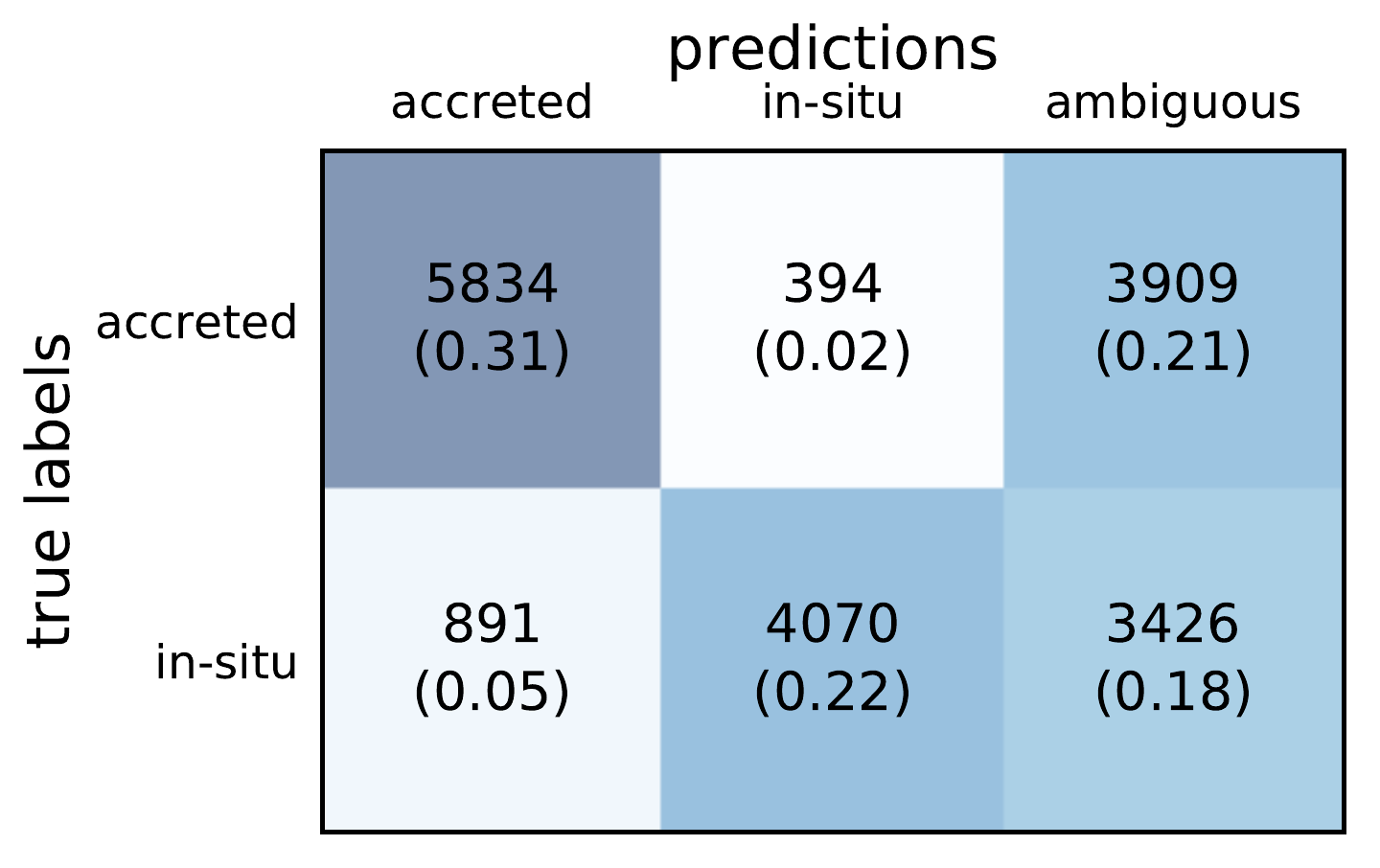}
    \caption{Confusion matrix showing the distribution of the predicted versus true labels of GCs in the test set. The `ambiguous' label corresponds to predictions with low confidence, $P<\Pthresh=0.79$. For each category the matrix shows the number of GCs, and the fraction relative to the total sample in parentheses. The background shading is darker for larger fractions. The model is excellent at classifying accreted GCs (with only 6 per cent falsely identified as in-situ), but misclassifies in-situ GCs in 18 per cent of the cases.}
    \label{fig:confusion_matrix}
\end{figure}

The results of the classification of the test set are shown in Figure~\ref{fig:confusion_matrix} using the standard confusion matrix. The columns represent the true labels, and the rows show the number of GCs in each column that are predicted to be in-situ, accreted, or ambiguous. After removing the ambiguous predictions, the model erroneously classifies 6 per cent ($394/5834$) of the accreted GCs, and a much larger fraction of in-situ GCs ($891/4070 = 18$ per cent).     

\begin{figure}
    \centering
    \includegraphics[width=\columnwidth]{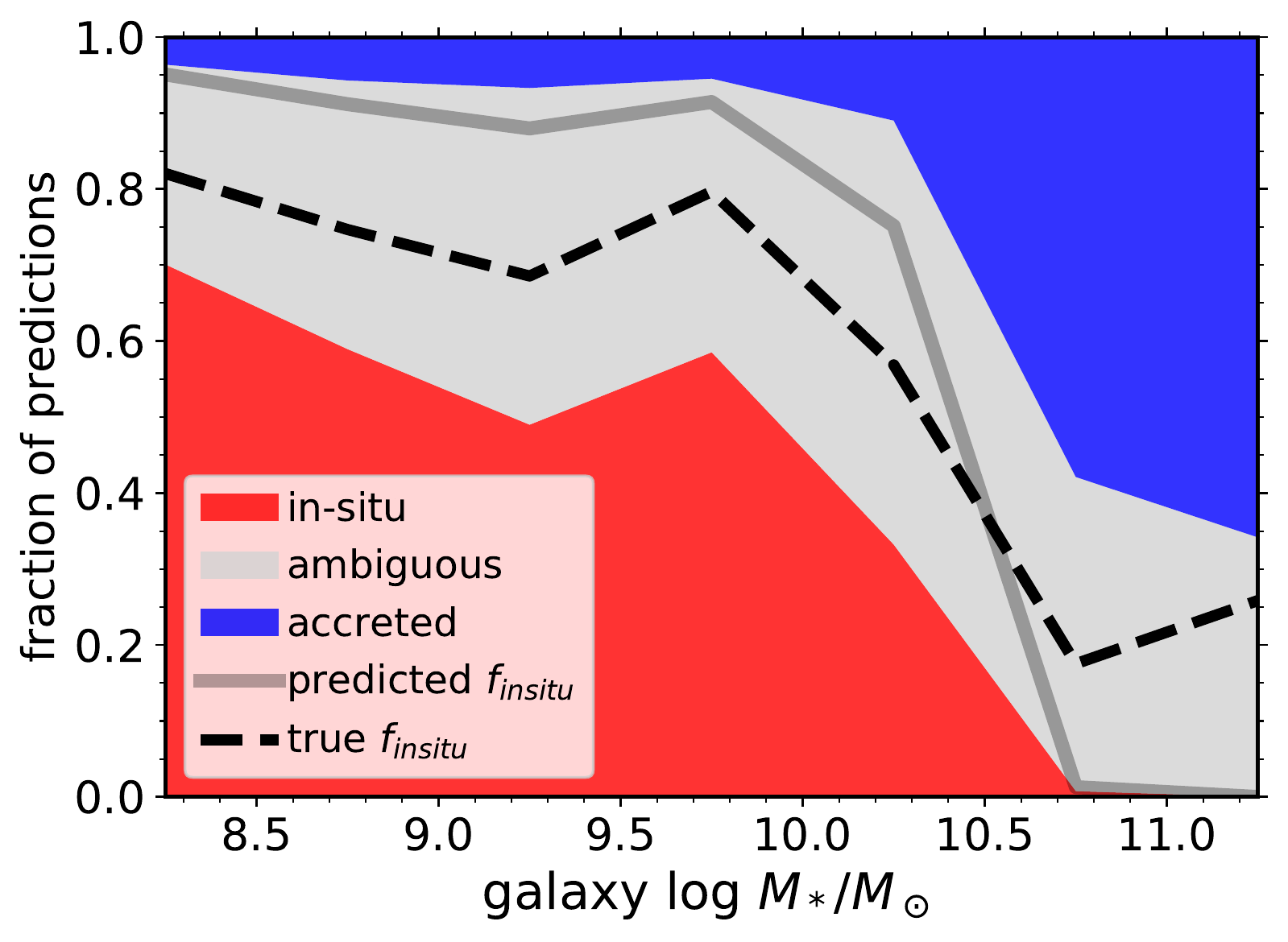}
    \caption{Fraction of test GCs in the predicted classes as a function of host galaxy stellar mass. The `ambiguous' class corresponds to predictions below the confidence threshold, $P<\Pthresh=0.79$. The grey line and grey shaded area show the predicted in-situ fraction and the uncertainty range (due to the ambiguous predictions). The dashed line indicates the true in-situ fraction. Even though $\sim25{-}45$ per cent of the sample is classified as ambiguous at a given mass, the model correctly predicts the majority class as a function of stellar mass.}
    \label{fig:class_fractions}
\end{figure}

The fraction of predicted labels for each class is shown in Figure~\ref{fig:class_fractions}. The ambiguous class represents a nearly constant fraction $\sim 25{-}45$ per cent of the predicted labels across the entire range of host galaxy stellar masses. The figure also shows that the model correctly predicts the dominant GC origin as a function of host galaxy stellar mass. Indeed, we find that the predictive accuracy remains nearly constant as a function of galaxy mass. However, the fraction of the dominant class is slightly overpredicted in dwarfs and massive ellipticals (but still lies within the uncertainty defined by the grey band).

\begin{figure*}
    \includegraphics[width=0.95\columnwidth]{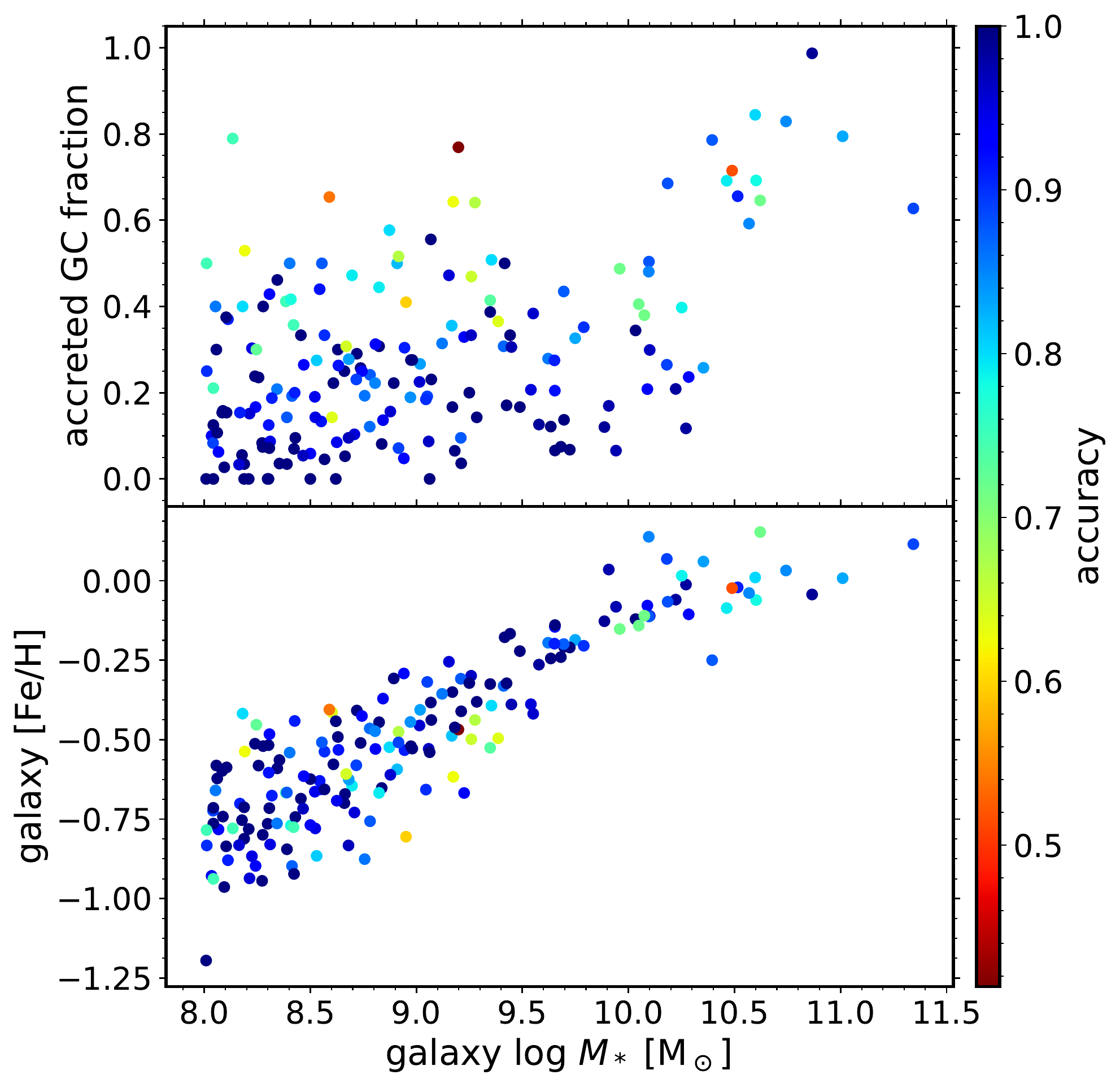}
    \includegraphics[width=0.95\columnwidth]{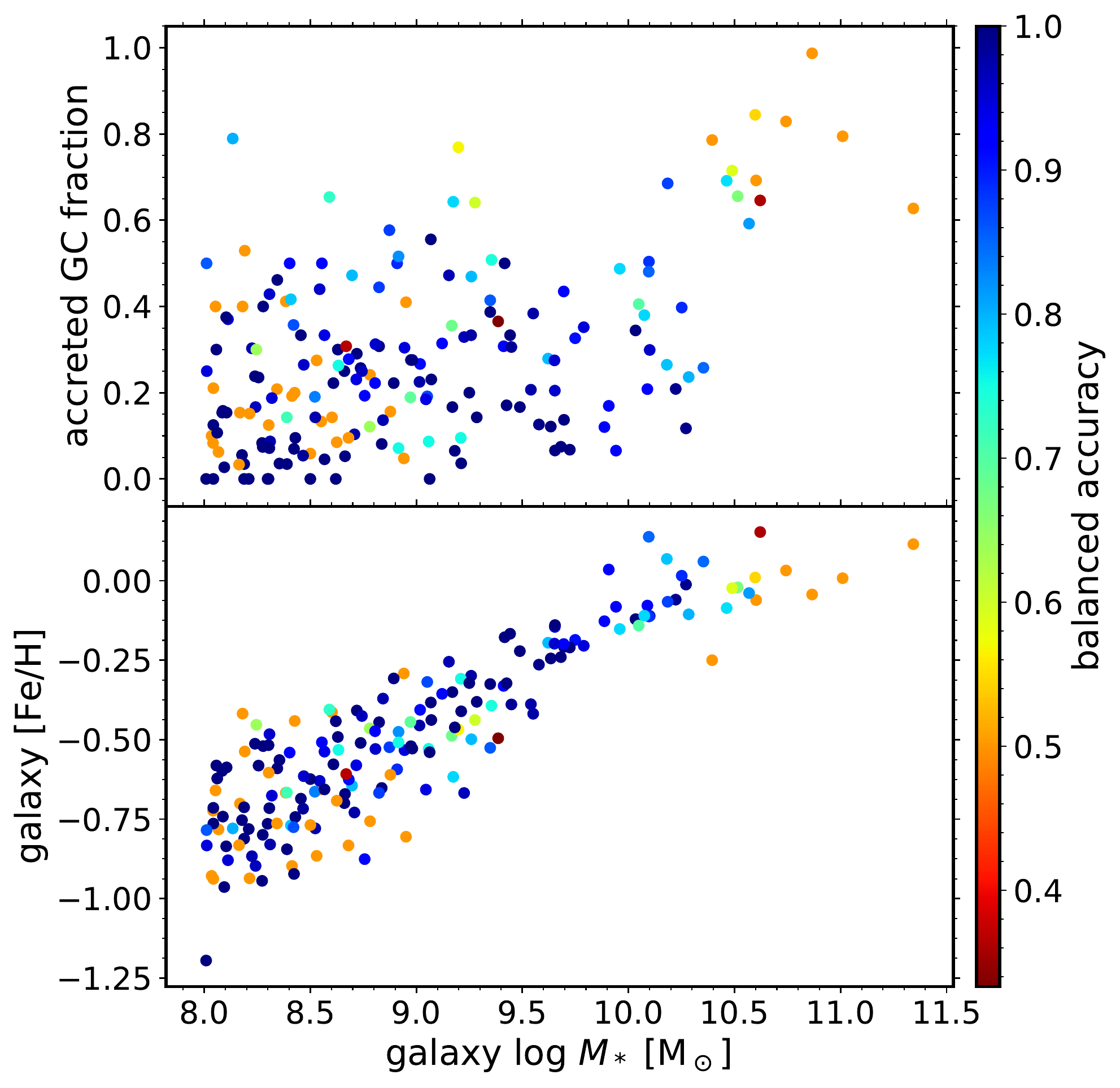}
    \caption{Predictive accuracy of the classifier on the simulated GC test set as a function of host galaxy properties. Left: accuracy across each galaxy versus galaxy mass and accreted GC fraction (top) and galaxy [Fe/H] (bottom). Right: same coloured by balanced accuracy. The accuracy tends to be lower in low-mass galaxies with high accreted fractions (the outliers in that mass range). The balanced accuracy in massive ellipticals is low because of the low performance of the model when predicting the origin of in-situ GCs (the minority class) as a result of the small number of training galaxies. The accuracy depends only weakly on galaxy metallicity.}
\label{fig:accuracy_galprops}
\end{figure*}

To evaluate the impact of global galaxy properties on the model performance, Figure~\ref{fig:accuracy_galprops} shows the accuracy obtained across each galaxy as a function of galaxy stellar mass, metallicity, and GC accreted fraction. To properly account for the large class imbalance in some galaxies (i.e. where accreted or in-situ GCs dominate), we also show the balanced accuracy (defined as the average of the accuracies calculated separately for each class). The accuracy reaches $>80$ per cent for the majority of galaxies, while it drops below 60 per cent in only a few galaxies. The low values of balanced accuracy in the most massive galaxies and several dwarfs are due to poor performance in identifying GCs in the minority class (which corresponds to in-situ for massive galaxies, and accreted in some dwarfs). This is more common among the most massive galaxies due to the small number of these objects in the training set. In addition, accreted GCs in massive galaxies have properties that are very similar to in-situ GCs due to the high masses of their satellite progenitors (see Sec.~\ref{sec:importances}). There is a weak trend of decreasing accuracy in metal-poor galaxies, which reflects the weak correlation between accreted fraction and metallicity (see Fig.~\ref{fig:accretedfrac_feh}).  

\subsection{Detailed predictions for simulated GC systems}

\begin{figure*}
    \includegraphics[width=0.65\textwidth]{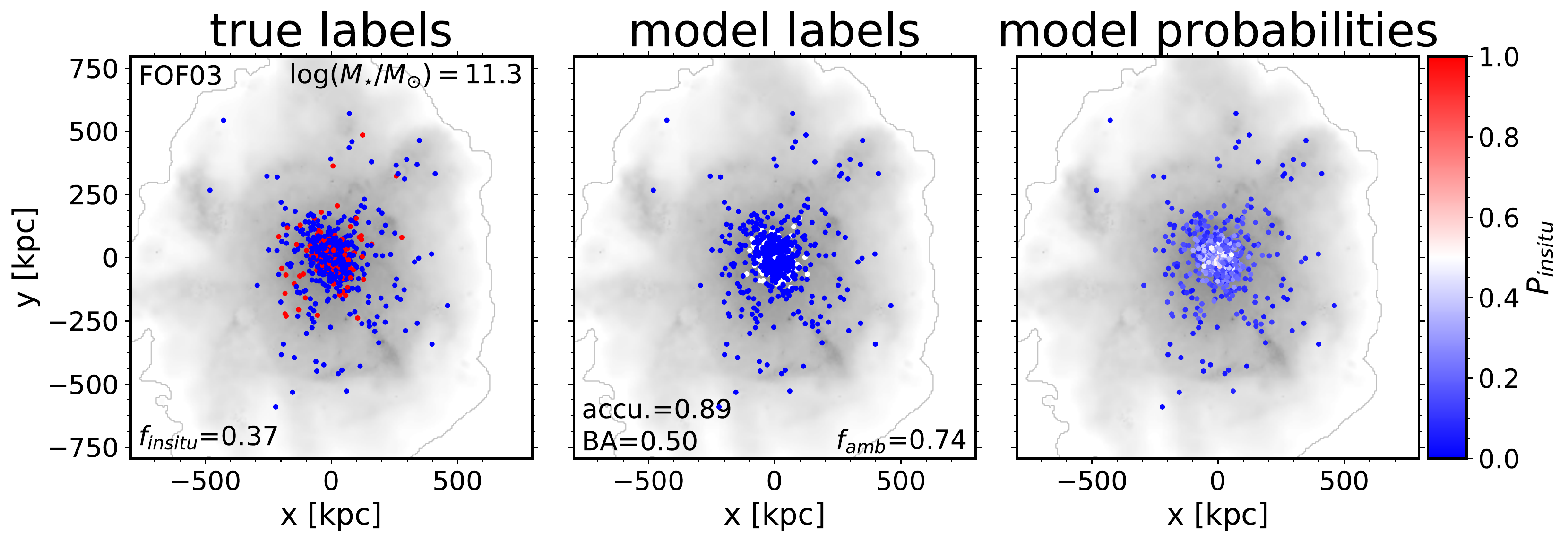}
    \includegraphics[width=0.65\textwidth]{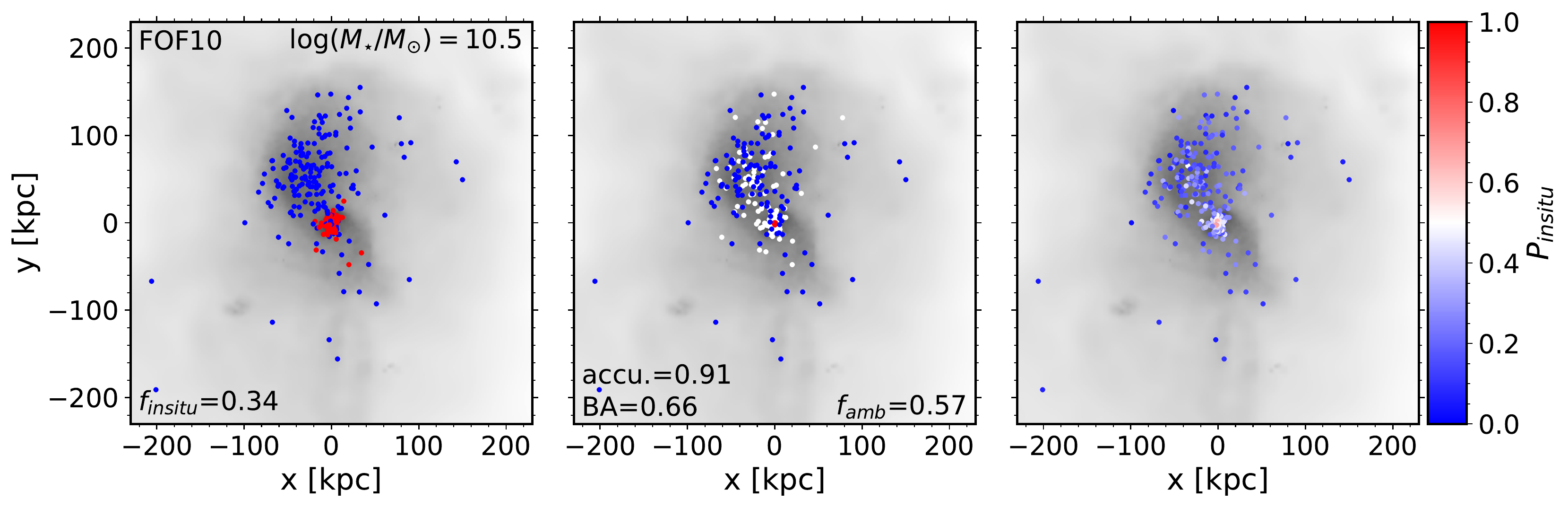}
    \includegraphics[width=0.65\textwidth]{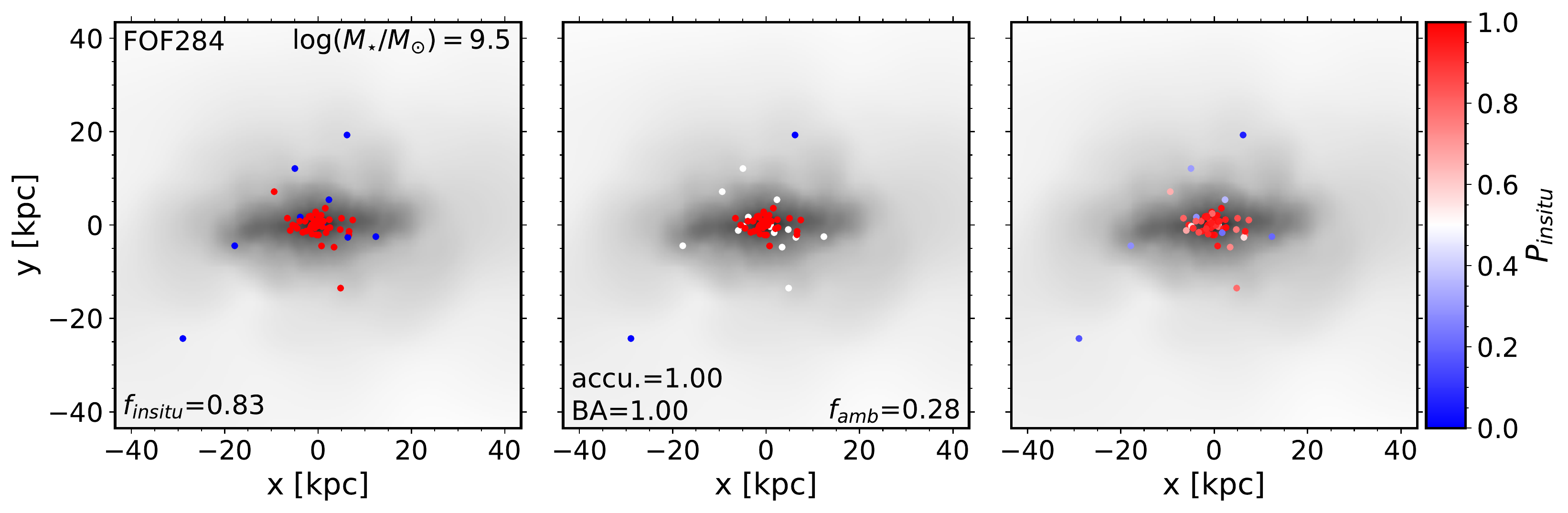}
    \includegraphics[width=0.65\textwidth]{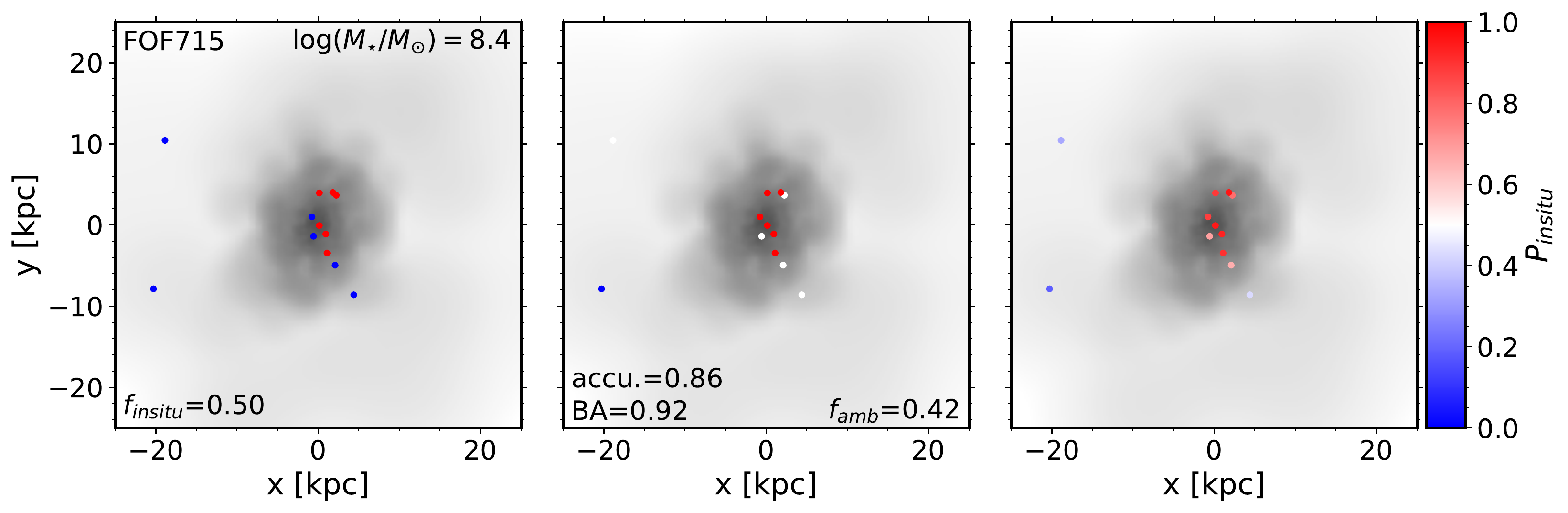}
    \caption{Projected distributions of GCs in the test sample and their predicted origin compared to their true origin. Each row shows the GCs hosted by selected galaxies (dots) in the random test set in each of four representative stellar mass bins, ranging from massive ellipticals (top row) to low-mass dwarfs (bottom row). The left column shows the true origin, while the middle and right columns show the predicted labels and probabilities of in-situ origin $\Pinsitu$, respectively. GCs with ambiguous classifications are shown in white in the middle column. The stellar mass, GC in-situ fraction, accuracy, balanced accuracy, and fraction of ambiguous predictions are indicated in each row. The stellar surface density is shown in grey-scale. The model produces high accuracy predictions for low-mass galaxies but has difficulty identifying in-situ GCs in massive galaxies with high accreted GC fractions due to their rarity in the training set.}
\label{fig:projections}
\end{figure*}

To evaluate the performance of the classifier across individual GCs, we now look at a few specific examples of galaxies in the test set. Figure~\ref{fig:projections} shows the projected distributions of GCs labeled by their predicted and true origin in four example galaxies selected randomly in each stellar mass bin, including a massive elliptical, a MW-mass galaxy, a massive dwarf, and a low-mass dwarf. In the massive elliptical galaxy, the model has difficulty identifying any of the in-situ GCs (as indicated by the low balanced accuracy), despite the galaxy containing 11 per cent of GCs in this class. Similarly, in the MW-mass galaxy that is currently undergoing a massive merger, the model achieves high accuracy overall but again has difficulty identifying the small fraction of in-situ GCs. Across the dwarf galaxies the model shows excellent performance on both classes (i.e. a high overall and balanced accuracy), despite the relatively low accreted fractions. The better performance in low-mass galaxies is consistent with their relative dominance across the training set. In general, the model seems to produce lower confidence predictions at intermediate galactocentric distances, where the in-situ and accreted GCs are co-spatial.  

\begin{figure*}
    \includegraphics[width=0.65\textwidth]{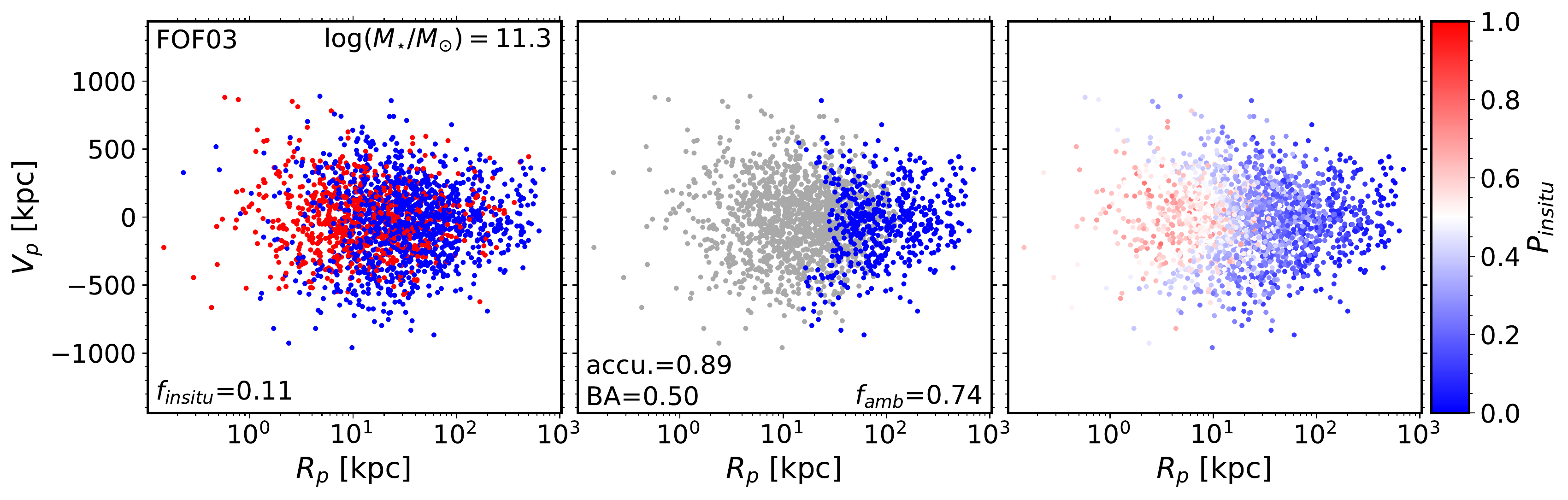}
    \includegraphics[width=0.65\textwidth]{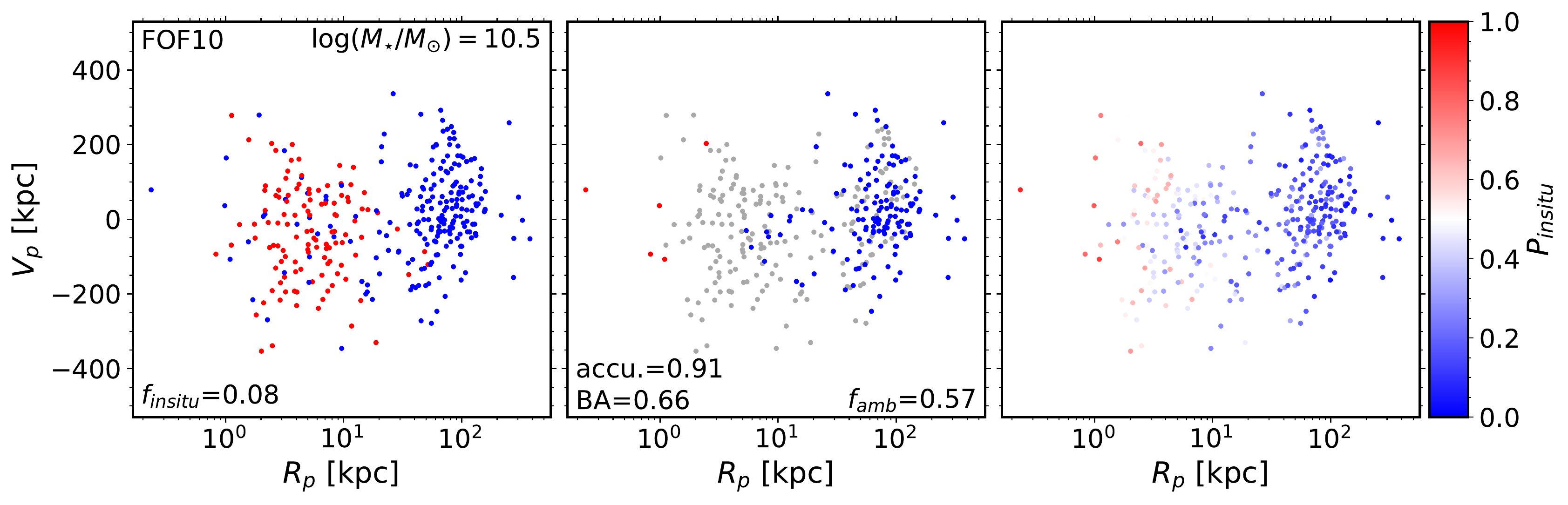}
    \includegraphics[width=0.65\textwidth]{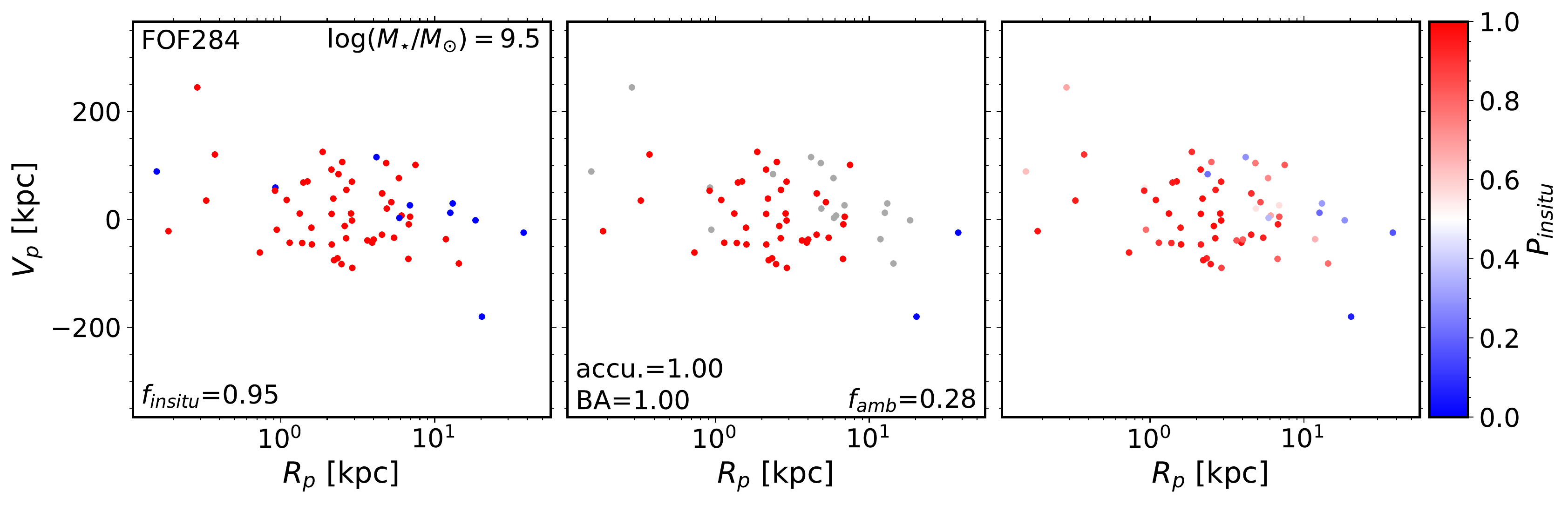}
    \includegraphics[width=0.65\textwidth]{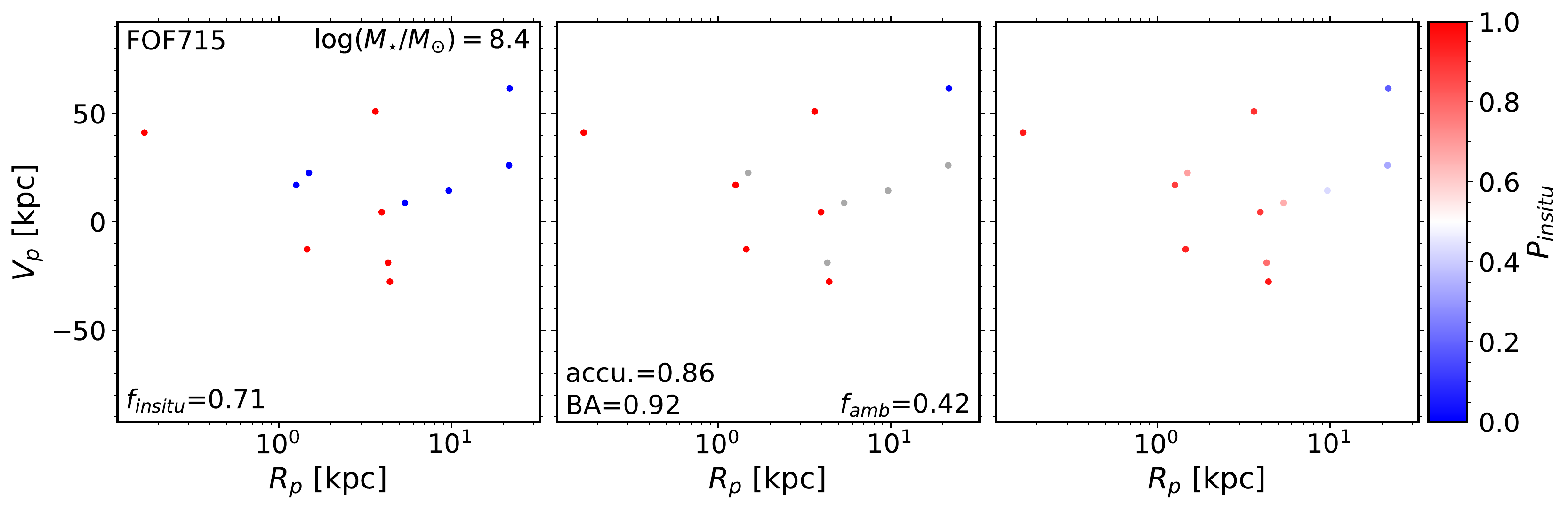}
    \caption{Projected position-velocity distributions of GCs in the test sample and their predicted origin compared to their true origin. The rows show the LOS velocity versus projected galactocentric radius for the randomly selected simulated galaxies in Fig.~\ref{fig:projections}. The left column shows the true origin, while the middle and right columns show the predicted labels and probabilities of in-situ origin $\Pinsitu$, respectively. GCs with ambiguous classifications are shown in grey in the middle column. The stellar mass, GC in-situ fraction, accuracy, balanced accuracy, and fraction of ambiguous predictions are indicated in each row. The decision boundary is clear in the right panels for massive galaxies, and highlights the predictive power of the `projected angular momentum' $\Rp|\Vp|$.}
\label{fig:projectionsRV}
\end{figure*}

To understand the role of the GC phase-space distribution in the model predictions we show the same galaxies in projected position-velocity space in Figure~\ref{fig:projectionsRV}. The importance of the `projected angular momentum' $\Rp|\Vp|$ is evident in the massive elliptical, with the decision boundary of the algorithm describing a near circle in position-velocity space, equivalent to a nearly constant projected angular momentum. This separation is also clear in lower mass galaxies, but in those cases the model predicts a more complicated boundary based on additional GC properties including their chemical abundances (see Fig.~\ref{fig:feature_importance}). We investigate which GC properties are most important for predicting GC origin in the next section.

\subsection{Importance of each GC and galaxy observable}
\label{sec:importances}

\begin{figure*}
    \includegraphics[width=\columnwidth]{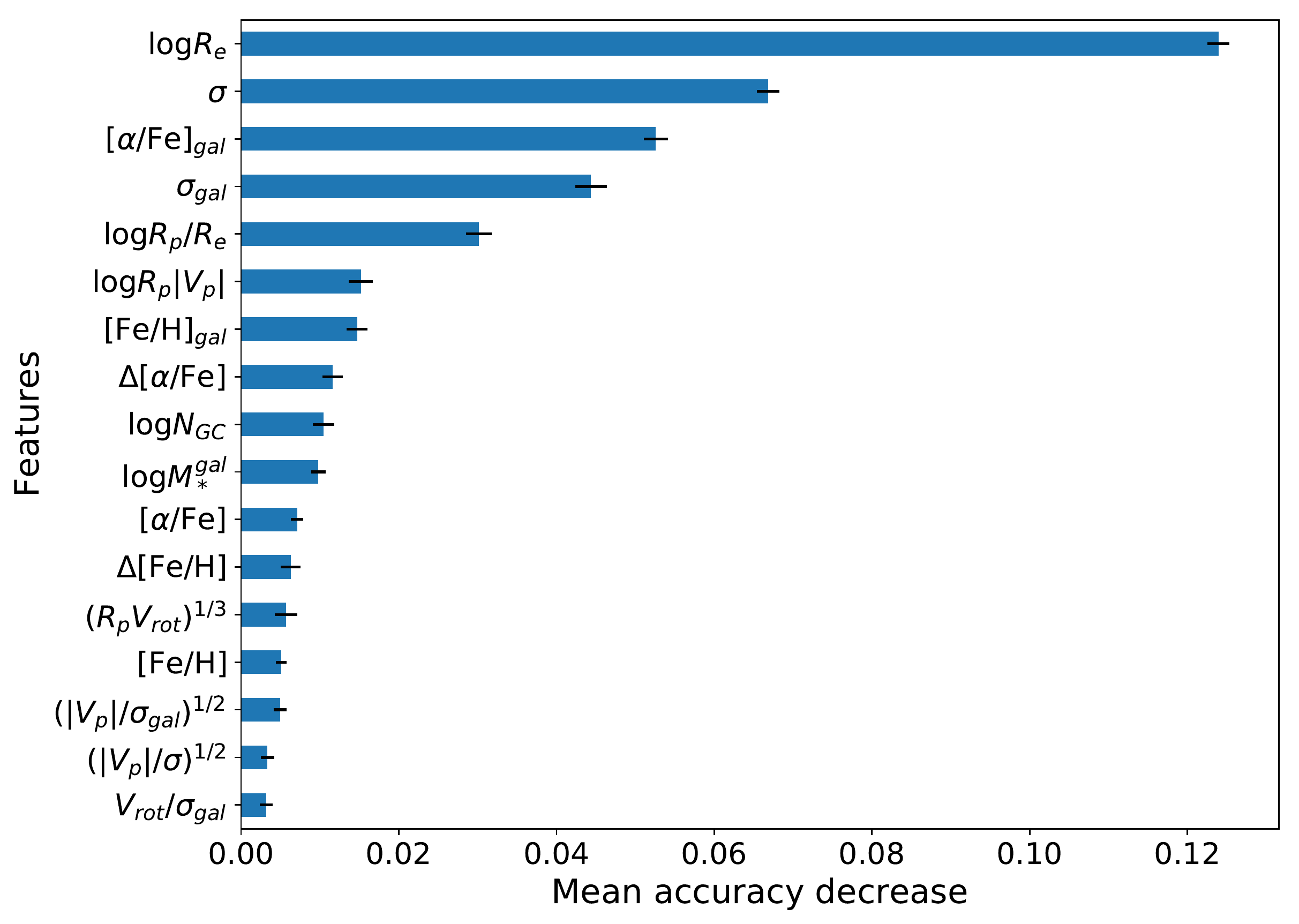}
    \includegraphics[width=\columnwidth]{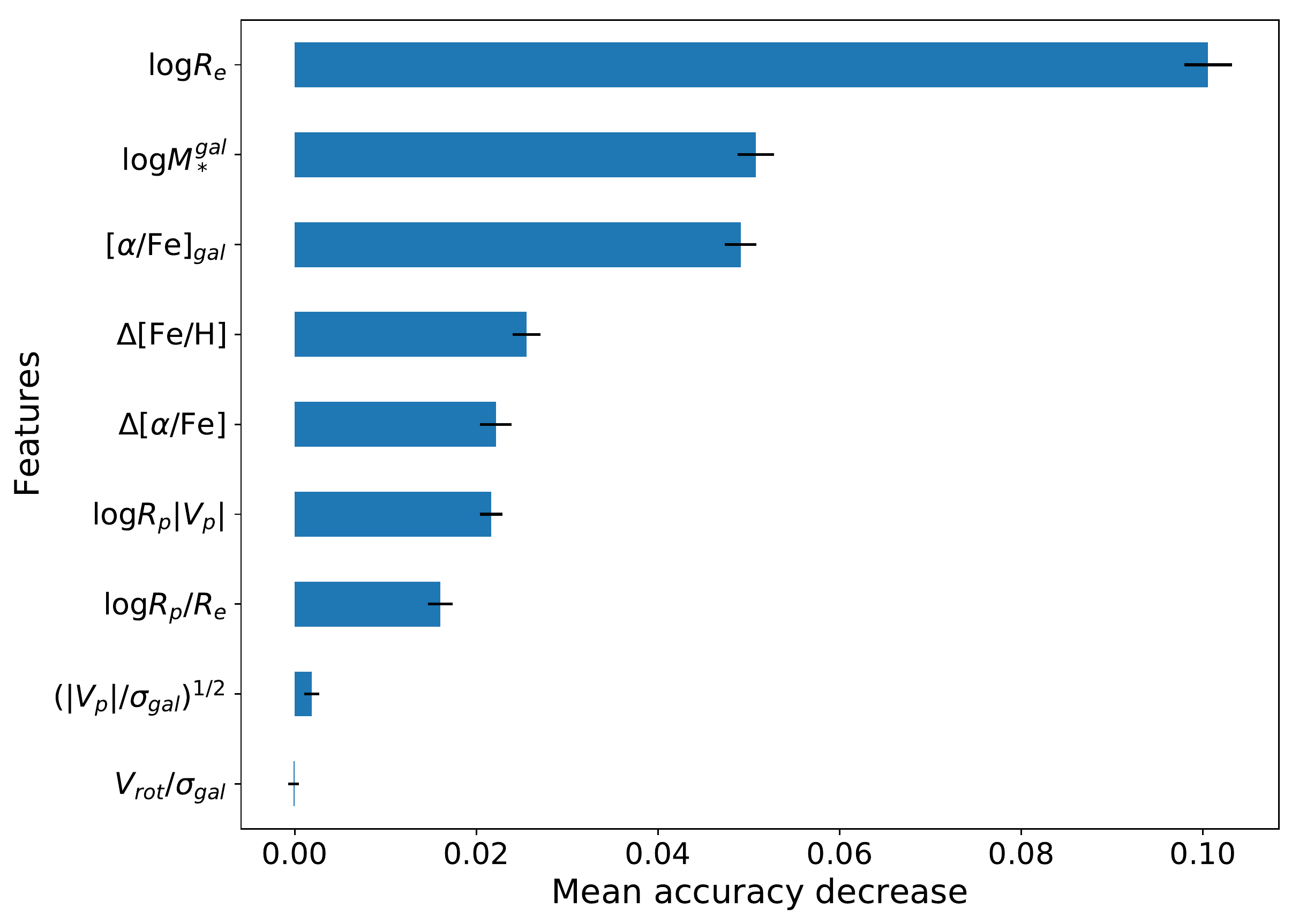}
    \caption{Permutation importance of each of the input features (i.e. observables) of the classifier. The value for each feature corresponds to the decrease in accuracy when the feature data in the test set is randomly shuffled before making predictions. Left: using all features. Right: After removing highly covariant features and retraining the model with only independent ones (see Sec.~\ref{sec:importances} for details). The black lines show the standard deviation in the result over 30 random iterations. The projected galaxy effective radius, stellar mass, and alpha-element abundance are the most predictive host galaxy properties. The most predictive GC observables are GC metallicity and alpha-abundance relative to the host galaxy, and projected angular momentum $\Rp|\Vp|$ and relative projected radius $\Rp/\Re$.}
\label{fig:feature_importance}
\end{figure*}

To assess how important each GC and galaxy observable is for the predictions of the model, we calculate the `permutation feature importances' \citep{Breiman01}. For a given feature, its importance is defined as the mean decrease in accuracy when the feature information in the test set is removed from the model input. For this, the feature vector of the desired feature is randomly shuffled while leaving the other features unchanged. The accuracy is then computed using the predictions over several random realizations of the shuffled data $\Niter$. The importance is then the difference in accuracy between the shuffled data and the fiducial model averaged over all realizations. The left panel of Figure~\ref{fig:feature_importance} shows the result using $\Niter=30$. 

Surprisingly, the most important features are host galaxy properties: the 2D effective radius, velocity dispersion (which is very similar for GCs and stars), and alpha-element abundance. These are followed by the GC projected galactocentric radius, the projected angular momentum, galaxy metallicity, and GC alpha-abundance offset relative to the galaxy. The importance of the galaxy properties might seem counterintuitive at first glance. However, it can be explained in two ways. First, most of the galaxy properties we use here correlate strongly with stellar mass, such that the classifier can obtain galaxy mass or size information indirectly from any of them. As we show in Fig.~\ref{fig:accretedfrac_Mhalo}, galaxy mass is the strongest predictor of GC accreted fraction, so it is natural for the algorithm to use it to estimate to first order the likelihood of a GC having formed in-situ. Second, highly covariant features can skew the results of the permutation technique, artificially reducing the importance of all features in a covariant cluster \citep{Wei15}. This occurs because the model can always obtain the information on a permuted feature from one of its covariates. 

To remove this possible bias, we perform a clustering analysis of all the features and split them into covariant groups based on a correlation threshold, and select only one feature from each group (see Appendix~\ref{sec:clustering_analysis} for details). We train a new model with the fiducial architecture but using only the selected subset of features\footnote{Since removing covariant features may lead to a slight loss of predictive power, we use this model variant only for the purpose of evaluating feature importance.}, and obtain the new permutation importances (right panel of Fig.~\ref{fig:feature_importance}). The three remaining galaxy properties still have the highest importance, followed by the GC metallicity and alpha-abundance relative to the galaxy, and the projected angular momentum and projected galactocentric radius in units of $\Re$. 

For galaxies in surveys with limited data (i.e. no alpha abundances), Figure \ref{fig:feature_importance} also provides an estimate of the performance if the model when a specific observable is not included. However, the optimal solution in this case would be to retrain a new model with the reduced feature set (see Section~\ref{sec:MW_test} for a discussion of the performance of such a reduced model).

\begin{figure}
    \includegraphics[width=1.0\columnwidth]{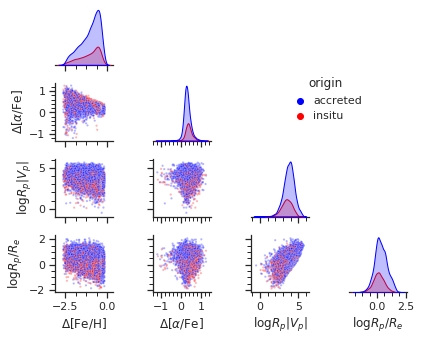}
    \caption{Joint and marginal distributions of GC origin across the observables with the most predictive power for GCs hosted by massive ellipticals. The panels show the distribution of in-situ and accreted GCs across simulated galaxies with $\Mstar>10^{11}\Msun$. The overlap of the two classes across all the observables partly explains the underperformance of the classififer in the most massive galaxies.}
\label{fig:corner_plots_ellipticals}
\end{figure}

In Appendix~\ref{sec:corner_plots} we show that the observables with the highest importances have  distributions across the GC sample that lead to the most distinct separation of in-situ and accreted objects. To understand why the classifier performs poorly in massive elliptical galaxies, Figure~\ref{fig:corner_plots_ellipticals} shows the distribution of GC observables for galaxies with $\Mstar>10^{11}\Msun$. The buildup of elliptical galaxies is dominated by massive satellites that contribute GCs with similar chemical abundances to the main progenitor GCs, and violent relaxation further mixes the two populations in phase space. The GC observables of in-situ and accreted populations entirely overlap, and this partly explains why the model cannot discriminate between the two classes from these data.

\subsection{Estimating uncertainty in the model predictions}
\label{sec:confidence}

Ideally we would like to predict not only the GC origin of each GC in an external galaxy, but also to have an idea of the uncertainty in the prediction. To estimate this predictive uncertainty we formulate a new problem: can we predict the accuracy of the model across a galaxy using only the observed properties of the galaxy? This would provide an estimate of how much the predictions for a given observed galaxy can be trusted. We explored a variety of regression algorithms including a Multilayer Perceptron with a linear activation function for the output layer \citep{Rumelhart86}, a Random Forest \citep{Breiman01}, and a Ridge Regressor \citep{Hoerl70}. Each model was trained on all the galaxy features listed in Table~\ref{tab:table2}, in addition to the features describing the distribution of GC galactocentric radii and LOS velocities in each galaxy (their mean, inter-quartile range, skewness, and kurtosis). 

\begin{figure}
    \includegraphics[width=1.0\columnwidth]{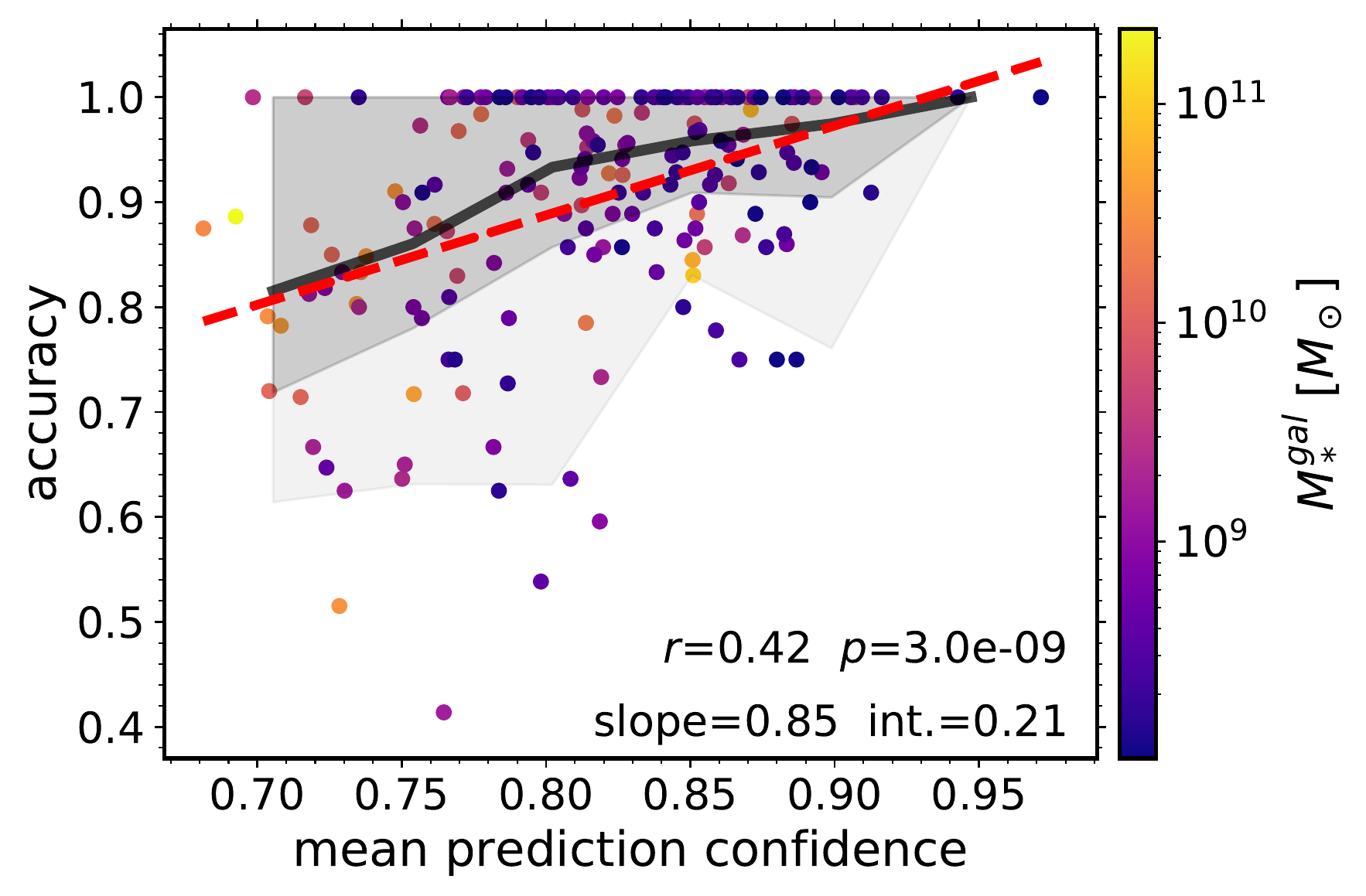}
    \caption{Accuracy of the ANN classifier as a function of the mean confidence in the predictions across each galaxy in the test set. The black line shows the binned median, and the dark and light shading contain the top 75 and 95 per cent of the accuracy distribution in each bin. The dashed line shows a linear fit, with parameters given in the legend. Here we define confidence as the maximum of the predicted class probabilities for each GC, $\max(\Pin,\Pacc)$. There is a highly significant correlation between mean prediction confidence and accuracy. About 75 per cent of simulated galaxies with a mean prediction confidence $\sim 0.85$ reach at least 90 per cent accuracy.}
\label{fig:accuracy_confidence}
\end{figure}

Perhaps unsurprisingly, we find that none of these algorithms can predict the accuracy of the fiducial classifier. To predict the galaxy-wide accuracy, the models would need to know the true GC origin labels, and this is precisely the information we lack for real galaxies. Deep learning offers a possible solution: the output of MLP classifiers is a set of class membership probabilities. We may therefore exploit the correlation that was found between the label probabilities $\Pin$ and the full sample accuracy in Fig.~\ref{fig:accuracy_Pinsitu} to predict the model uncertainty. We define the `confidence' of the model predictions for a galaxy by how close on average the predicted probabilities get to complete certainty,
\begin{equation}
    {\rm mean ~confidence} = \frac{1}{\Ngc}\sum_{i=1}^{\Ngc}\max\left(\Pin^i,\Pacc^i \right) .
\end{equation}
We examine the relation between the galaxy-wide accuracy and mean prediction confidence using the simulation test set in Figure~\ref{fig:accuracy_confidence}. To calculate the mean confidence we use all the GC predictions, including those with $P<\Pthresh$. Despite the large scatter, there is a highly significant correlation ($p=3\times10^{-9}$) between mean prediction confidence and accuracy. The median accuracy increases from $\sim 0.8$ to $\sim 1.0$ as the mean confidence increases from $\sim 0.70$ to $\sim 0.95$. This shows that the neural network successfully learned which regions of the high-dimensional feature space contain both in-situ and accreted GCs, and therefore lead to ambiguous predictions. We can then use the distribution of galaxy-wide accuracy in Fig.~\ref{fig:accuracy_confidence} to estimate the probability that the classifier will reach a given desired accuracy in a real galaxy. For instance, we expect that the classifier will be more than $90$ per cent accurate in 3 out of 4 galaxies that reach a mean confidence $\sim 0.85$. The dashed line in Fig.~\ref{fig:accuracy_confidence} shows a linear fit to the data with the parameters provided in the legend.   

\subsection{Testing the model on the Milky Way GCs}
\label{sec:MW_test}

Simulations are rough simplifications of the real Universe. As such, they may or may not capture the physical processes linking GC formation to their observable properties. With any supervised deep learning model trained on simulation data the question therefore arises: does the complex relationship between the features and target variable learned by the model resemble the actual relation in the real Universe? In other words, does the performance of the model using real data match the performance on the simulated test data? To answer this question we now perform a first, real-world test of the ANN classifier using data for the Milky Way GCs.

For this test we use the detailed data on the MW GC system that has been compiled over several decades, together with the progenitor associations determined recently using \emph{Gaia} orbital information, chemical abundances, and ages \citep{Massari19,krakenI,krakenII}. We use the compilation of GC metallicity data from \citet[][2010 edition]{Harris96}, and the 3D positions and velocities compiled by \citet{Baumgardt19} from a combination of \emph{HST} and \emph{Gaia} data. To extend the applicability of the model to surveys that do not include the most difficult to obtain GC observables, we build a new `minimal' ANN classifier using a reduced feature set (by removing the alpha-element abundances and velocity dispersions), and train it using the fiducial simulation training set. As we show below, this retraining procedure achieves a better perfomance than simply neglecting these features in the fiducial model (where the loss of accuracy would be $> 5$ per cent, see Fig. \ref{fig:feature_importance}). Table~\ref{tab:table3} summarizes the features of the minimal classifier.

\begin{table*}
\centering{
  \caption{GC and host galaxy observables used as features in the `minimal' classifier. Projected positions and LOS velocities are calculated with respect to the position and velocity of the centre of the galaxy, assuming a single random orientation for each galaxy.}
  \label{tab:table3}
	\begin{tabular}{lcl}\hline\hline
		Feature & Object & Definition  \\ 
		\hline
		$\log\Mstargal$ & galaxy & stellar mass \\
		$\log\Regal$ & galaxy & projected effective radius \\
		$\fehgal$ & galaxy & mean metallicity \\
		$\fehgc$ & GC & metallicity \\
		$\deltafeh$ & GC/galaxy & metallicity relative to the galaxy, $\fehgc - \fehgal$ \\
		$\log\Rp/\Regal$ & GC/galaxy &  projected distance from galaxy centre in units of the galaxy effective radius \\
		$\log\Rp|\Vp|$ & GC & `projected angular momentum': product of projected galactocentric distance and LOS velocity \\
		\hline \hline
	\end{tabular}}
\end{table*}

For the global properties of the Galaxy we assume $\Mstargal = 5\times10^{10}\Msun$, $\Regal = 3.8\kpc$ \citep{Cautun2020}, and $\fehgal = 0.0$ \citep{Bland-HawthornGerhard16}. We apply the same metallicity selection used for the simulation to the MW GCs (see Table~\ref{tab:table1}), without imposing a GC mass cut (since this was only used to remove artefacts in the simulation). This results in a sample of 129 GCs with $-2.5<\feh<-0.5$. For the true origin labels we use the classification by \citet{Massari19} (as revised by \citealt{krakenII} for Pal 1 and NGC 6441) based on the GC ages, metallicities, and orbits. To obtain the projected positions and LOS velocities we artificially incline the plane of the Galaxy by an angle $i \deg$ (around the x-axis) towards the observer. 

As in the case of the fiducial model, we optimize the architecture using a grid search for the combination $[\Nlayers,\Nnodes]$ that yields the highest accuracy on the simulation test set. For a decision threshold $\Pthresh=0.5$, the resulting network achieves an accuracy of $\sim 78$ per cent on the test data (using $\Nlayers=2$ and $\Nnodes=50$). This corresponds to a decrease of $\sim 2$ per cent compared to the fiducial model. During testing of the minimal model we found that the accuracy of the MW predictions varies significantly across identically trained models (with a dispersion of $\approx 3$ per cent) as a result of the inherent stochasticity in the ANN training process\footnote{This is a well known trade-off of the computational efficiency necessary for estimating the gradient of the loss function in a high-dimensional feature space.}. This stochasticity is averaged out when considering the large simulated GC test sample, but becomes more important when evaluating the predictions for the small set of GCs in the MW system (see Appendix~\ref{sec:minimal_model}).

\begin{figure*}
    \includegraphics[width=\columnwidth]{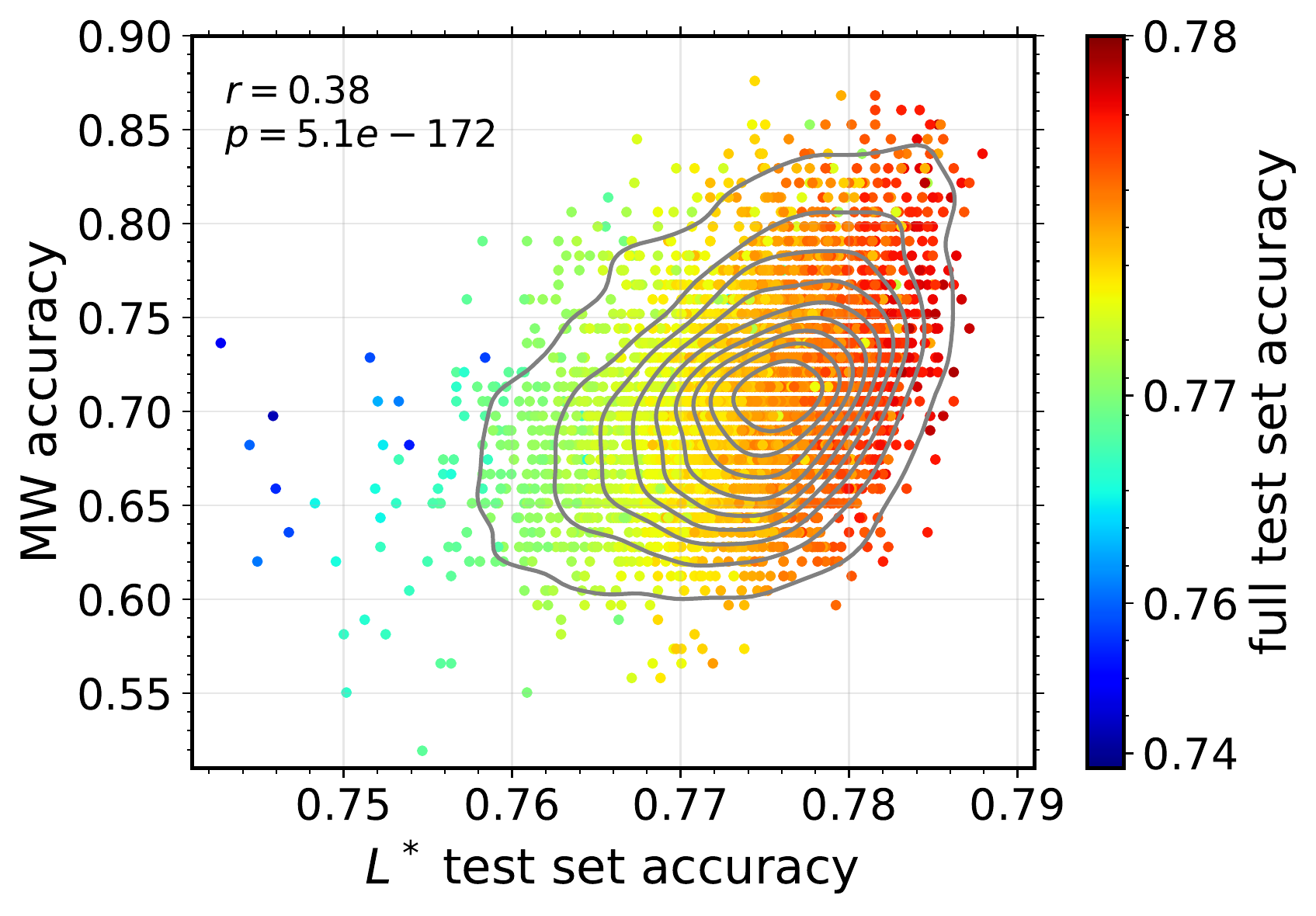}
    \includegraphics[width=0.95\columnwidth]{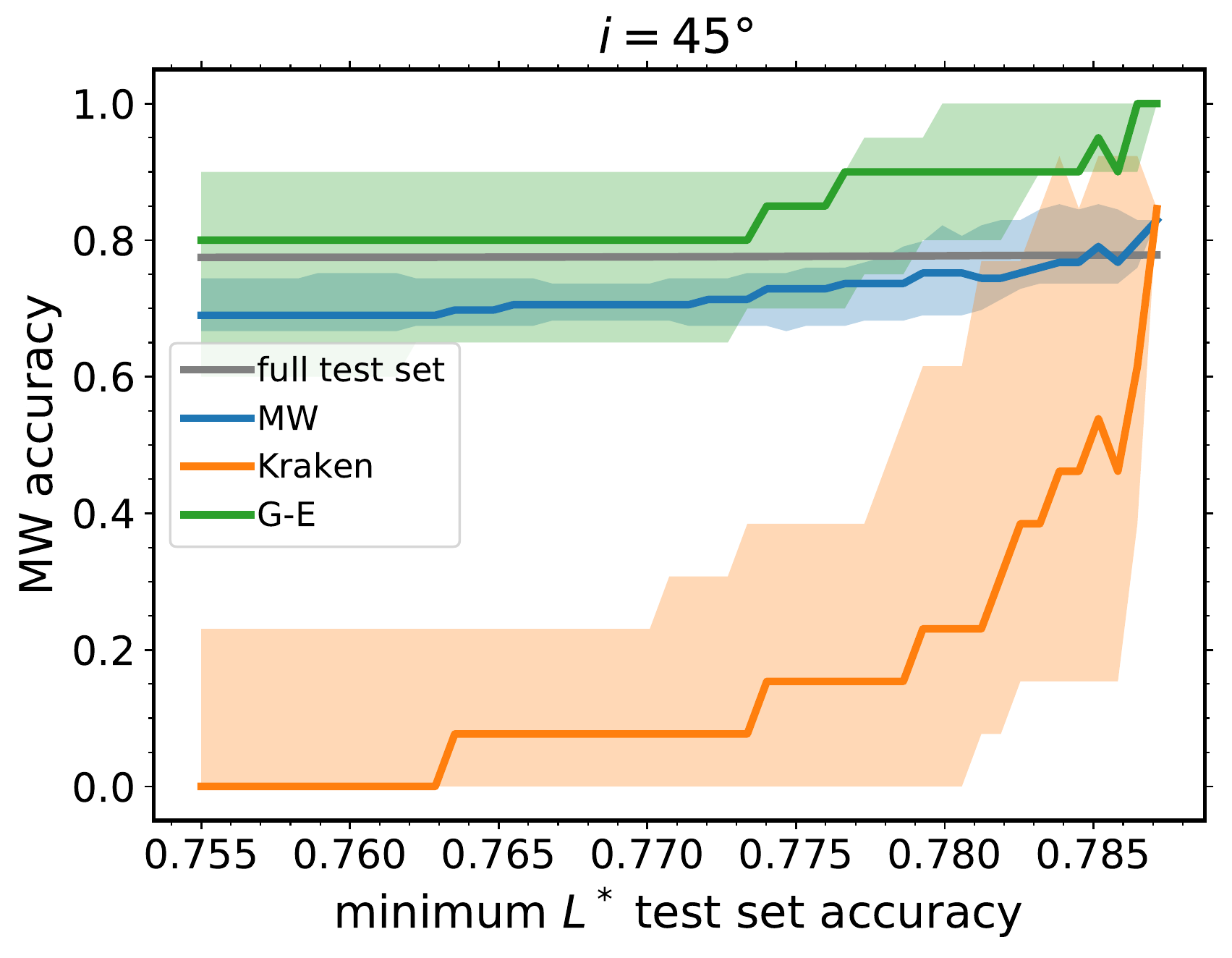}
    \caption{Performance of an ensemble of minimal classifiers on the simulated test set and on the MW GCs. Left: correlation between the accuracy of each model (points) on the MW GCs and on the 24 simulated $L^*$ galaxies in the test set (and its Spearman coefficient and $p$-value), with the colour indicating the accuracy on the full test set, and the contours showing a kernel density estimate of the underlying distribution. Right: accuracy of a voting ensemble as a function of the minimum $L^*$ galaxy test set accuracy used in the selection. The ensemble uses 5000 models trained on identical simulation data (and model architectures sampled from a grid of $[\Nnodes,\Nlayers]$). To obtain the MW accuracy the models are tested on randomly inclined MW GC system observables. We exploit the strong correlation between test set accuracy and MW accuracy to select a model with the highest performance on both simulated and observed galaxies.}
\label{fig:MW_modelsample}
\end{figure*}

To reduce the variance in the MW predictions we create an ensemble of 5000 models trained on identical simulation data, and vary the network complexity by sampling uniformly from the grid of $[\Nnodes,\Nlayers]$ described in Sec.~\ref{sec:training}. We then tested each model on three different samples: the full simulation test set, the subset of 24 $L^*$ galaxies (i.e. $10^{10} \leq \Mstar/\Msun \leq 10^{11}$) in the simulation test set, and the projected MW GC system (at random inclinations sampled uniformly from the range $0\leq \cos i \leq 1$). The results are shown in Figure~\ref{fig:MW_modelsample} as a function of the performance on the $L^*$ galaxy test set. We find an interesting statistically-significant correlation between the performance on the $L^*$ simulations and on the real Milky Way, and a weaker correlation with the fulll test set. This indicates that models with above-average performance on the simulations will in general also produce more accurate predictions on real galaxies. In other words, the best models tend to have the best generalization capacity, and this would only be true if the simulation captures the physical processes responsible for the formation and evolution of the Galaxy. 

The perfomance of the model ensemble on each sample as a function of the accuracy threshold is shown in the right panel of Figure~\ref{fig:MW_modelsample}. In addition to the MW, we show the accuracy on the two main progenitors, \emph{Kraken} and \emph{Gaia-Enceladus}. The predictive accuracy of the ensemble increases with the threshold for the $L^*$ test sample, the MW, and its two main progenitors (and increases slightly for the full test set).  

\begin{figure*}
    \includegraphics[width=0.8\textwidth]{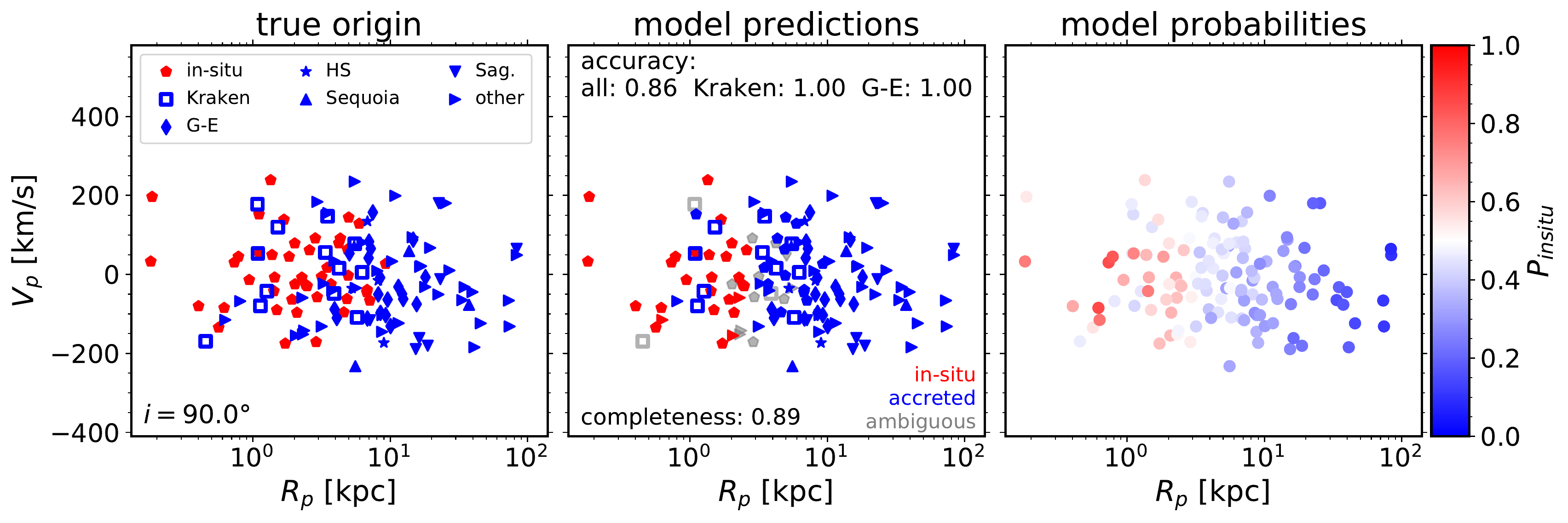}
    \includegraphics[width=0.8\textwidth]{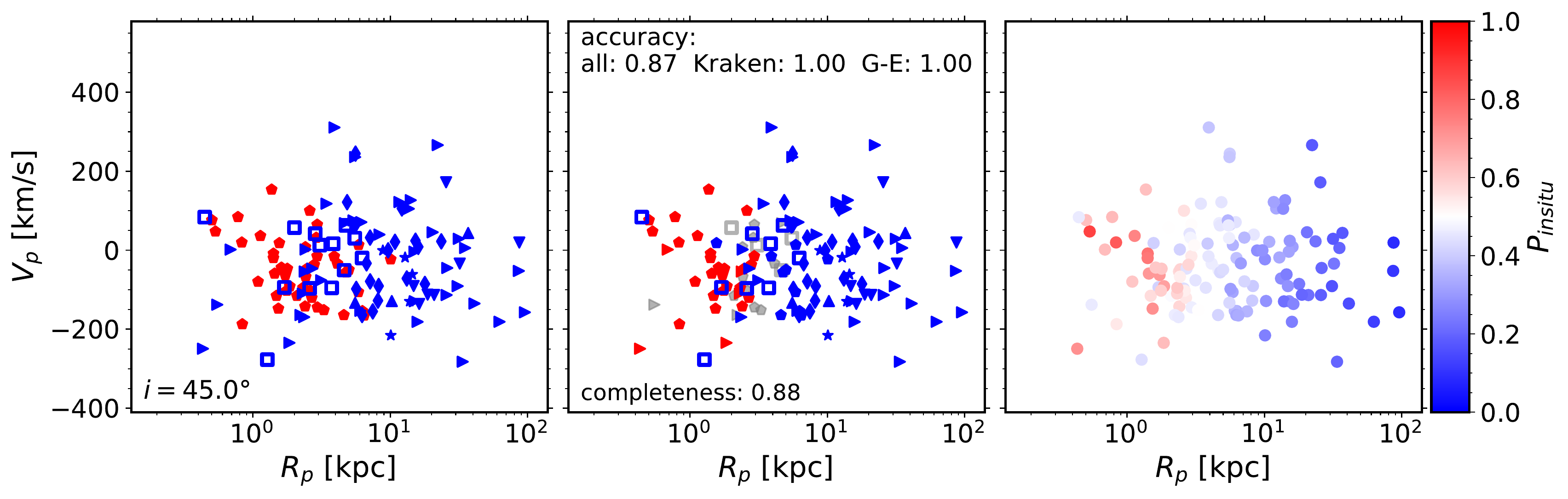}
    \includegraphics[width=0.8\textwidth]{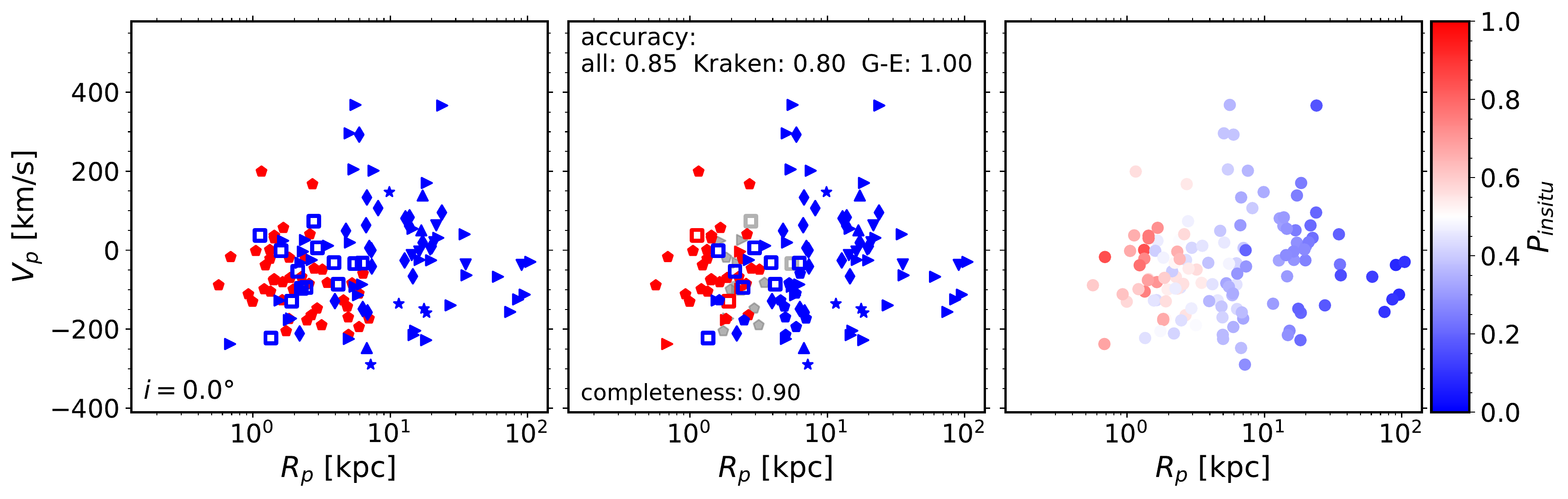}
    \caption{Predictions for the origin of the Milky Way GCs as a function of line-of-sight velocity and projected galactocentric distance. Each row shows the results assuming that the MW is observed at a different inclination, as indicated in the legend. Left: true origin is indicated using different symbols for each progenitor galaxy, and colors indicating in-situ and accreted GCs. Middle: predicted in-situ and accreted labels are indicated with using color, with ambiguous classifications shown in grey (and symbols corresponding to each progenitor). Right: predicted probability of in-situ origin $\Pin$. The accuracy for the entire GC system and for each of the two major progenitors is indicated in the middle panels. The performance of the minimal ANN classifier is robust to the assumed inclination of the Galaxy, and the model successfully identifies GCs in each of the five known progenitors, including at least $80$ per cent of the GCs associated with \emph{Kraken} (squares), the progenitor debris located closest to the centre of the Milky Way, and all of the \emph{Gaia-Enceladus} GCs. }
\label{fig:MW_RVplots}
\end{figure*}

To visualize the predictions, the first column of Figure~\ref{fig:MW_RVplots} shows the position-velocity diagram of the projected MW GCs coloured by their true origin, where each row corresponds to a different viewing angle. The other two columns show the origin labels (middle) and probabilities (right) predicted by the minimal classifier. To obtain the predictions we selected the model with the highest performance on the $L^*$ test set, and further optimized its performance by tuning $\Pthresh$ to achieve a high accuracy and low ambiguous fraction (see Appendix~\ref{sec:minimal_model} for details). Using only a single model from the ensemble may increase stochasticity (i.e. noise) in the results, but we checked explicitly that this is not the case when comparing to a voting ensemble of the 100 best models. Figure~\ref{fig:MW_modelsample} shows that selecting the model with the best performance on the $L^*$ simulation test set guarantees a high accuracy on the MW system (blue line), without sacrificing the performance (i.e.  due to overfitting) across the broad galaxy population (gray line).  

The best-performing model predicts the origin of up to $\sim 9/10$ of the MW GCs unambiguously with an accuracy of $85-87$ per cent overall, and $\geq 80$ and $100$ per cent for the \emph{Kraken} and \emph{Gaia-Enceladus} GCs respectively (adopting $\Pthresh=0.52$). Increasing the decision threshold to $\Pthresh=0.6$ improves the MW accuracy to $90$ per cent and the ambiguous fraction to 0.4 (see Appendix\ref{sec:minimal_model}). For the baseline value $\Pthresh=0.5$ the performance is comparable to the accuracy obtained on the full test set drawn from the simulations (grey line in the right panel of Fig.~\ref{fig:MW_modelsample}), and on the 24 simulated galaxies with masses $10^{10}<\Mstar<10^{11}\Msun$ (x-axis of right panel of Fig.~\ref{fig:MW_modelsample}).  The excellent performance on the MW GCs implies that the simulation training data accurately follows the physical processes that shape the observed properties of in-situ and accreted GC populations and their host galaxies in the real Universe, and that the ANN effectively learned this relation. 

The first column of Figure~\ref{fig:MW_RVplots} also indicates the known galactic progenitors associated to each GC (from \citealt{krakenII}) using different symbols. Out of the five known progenitors that contributed accreted GCs, only the \emph{Kraken} debris is located in the inner Galaxy, at galactocentric distances $r \la 10\kpc$. This could potentially make the classification more challenging for the model, since it relies partly on the projected galactocentric distance (see Section~\ref{sec:importances}). Despite this, we find that the model correctly identifies as accreted $8{-}10$ out of the 13 known \emph{Kraken} GCs (shown as squares), in addition to all the \emph{Gaia-Enceladus} GCs, and at least a few GCs in each of the other three progenitors. 

The success of the deep learning classifier in identifying debris from all the known MW progenitors has important implications for the observational reconstruction of the assembly histories of other galaxies, where only limited GC phase-space information is available. The accurate identification of accreted GCs by the model in this test shows that there is enough archaelogical information in extragalactic GC observables to partially reconstruct the merger trees of galaxies in large surveys. We will investigate this intriguing possibility in future work.

\subsection{Impact of uncertainties in observational data}
\label{sec:uncertainties}

The simulated observables used in training and evaluating the model so far assume measurements with perfect precision. Some GC and galaxy observables can include large uncertainties that arise either from the quality of the data, or from the methods used to infer the physical property from either the photometry or the spectra. Here we perform an analysis of the effect of uncertainties on the predictions to understand the sensitivity of the model, and to provide benchmarks for the expected behaviour of the model for given values of the uncertainties.

For this, we perform a Monte Carlo experiment. We first inject random noise following a normal distribution with the width given by the relative uncertainty in each of the GC observables in the test set, $\fehgc$, $\alphagc$, $\Rp$, and $\Vp$, in addition to the host galaxy properties $\log \Mstargal$ and $\log \Regal$. The position and velocity errors are calculated relative to the effective radius and velocity dispersion of the galaxy, respectively. We then use the fiducial ANN classifier (trained on the unperturbed data) to obtain predictions for uncertainties in the range $0.0{-}0.5$, equivalent to relative errors of up to a factor of 3 in stellar mass and effective radius, and absolute errors of up to 50 per cent of the effective radius and velocity dispersion for $\Rp$ and $\Vp$, respectively. 

\begin{figure}
    \includegraphics[width=1.0\columnwidth]{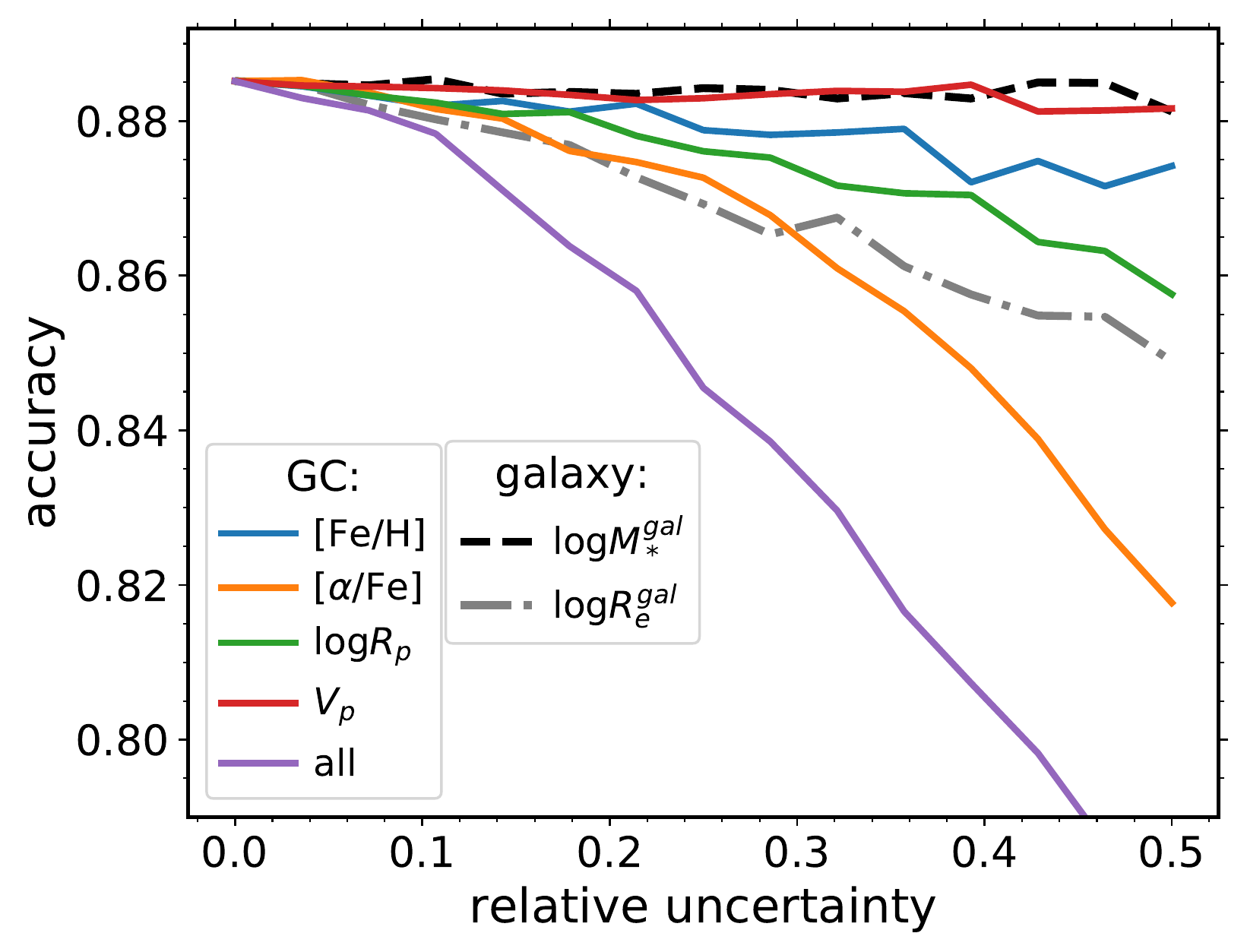}
    \caption{Impact of observational uncertainties on the accuracy of the GC origin predictions. Each line shows the accuracy as a function of the relative error in each of the GC observables: metallicity, alpha abundance, projected position, and line-of-sight velocity, in addition to the galaxy stellar mass and effective radius. The bottom line shows the effect of uncertainties in all the observables combined. For $\Vp$ the $x$-axis represents fractional uncertainty with respect to the velocity dispersion of the host galaxy, while for all other quantities it represents order of magnitude uncertainties (i.e. in dex). The values are obtained by adding normally-distributed Monte Carlo errors to the test set drawn from the simulations. The accuracy is robust to relative uncertainties as large as $\sim 0.1$ in the GC observables (see Section~\ref{sec:uncertainties} for the interpretation for each observable). The performance of the classifier is most sensitive to the precision of the alpha abundances and galaxy effective radius.}
\label{fig:accuracy_vs_obserrors}
\end{figure}

The resulting accuracy as a function of the relative observational uncertainty in each feature is shown in Figure~\ref{fig:accuracy_vs_obserrors} for each of the GC observables used in the fiducial feature set (see Table~\ref{tab:table2}). The uncertainty in the alpha-element abundances dominates the prediction errors, but still only produces a slight $\la 1$ per cent decrease in accuracy for errors as large as $0.2$ dex. The model is robust to large individual uncertainties in the GC projected galactocentric radius, LOS velocity, and metallicity. Including uncertainties in all the observables results in a $\sim 1.5$ per cent drop in accuracy for relative errors $\sim 0.15$. The observational uncertainties in distances and LOS velocities of extragalactic GCs are typically smaller. Distances of galaxies within $\sim 40\Mpc$ can be determined to $\sim 10$ per cent precision \citep[e.g.][]{Tonry01,Blakeslee09}, and velocities to a precision $\la 15\kms$ \citep[e.g.][]{Forbes17}, or about $\sim 12$ per cent of the MW velocity dispersion. Uncertainties in metallicity determinations are larger, $\sim 0.15$ dex \citep[e.g.][]{CaldwellRomanowsky16}, but still in the range where they would have a minimal effect on the accuracy of the model predictions. This analysis indicates that the uncertainties in the effective radius and the GC alpha abundances will be the dominant observational sources of error in the model's predictions.  

\subsection{Including GC ages to improve performance}
\label{sec:ages}

Due to limitations in the modelling of integrated spectra, GC ages are notoriously difficult to constrain beyond the Local Group \citep{Worthey94}. However, recent studies suggest that the precision of extragalactic GC ages can be improved significantly, reaching $\la 0.1$ dex relative uncertainties \citep{Usher19,Cabrera-Ziri22}. High precision GC ages in the local Universe could therefore be within reach for wide spectroscopic surveys over the next decade. In this section we test the effect of including the ages of the simulated GCs in training the ANN classifier. For this we add the precise GC age as an additional feature, and then evaluate the performance of the model on the test data from the simulation. We then run a Monte Carlo experiment to add random log-normal noise to the ages in the test data, and calculate the accuracy as a function of the uncertainty in the ages as well as in each of the other observables. As in the fiducial model, to remove ambiguous results we assume a decision threshold that predicts GC origin for $\sim 60$ per cent of the test sample, $\Pthresh=0.83$. 

\begin{figure}
    \includegraphics[width=1.0\columnwidth]{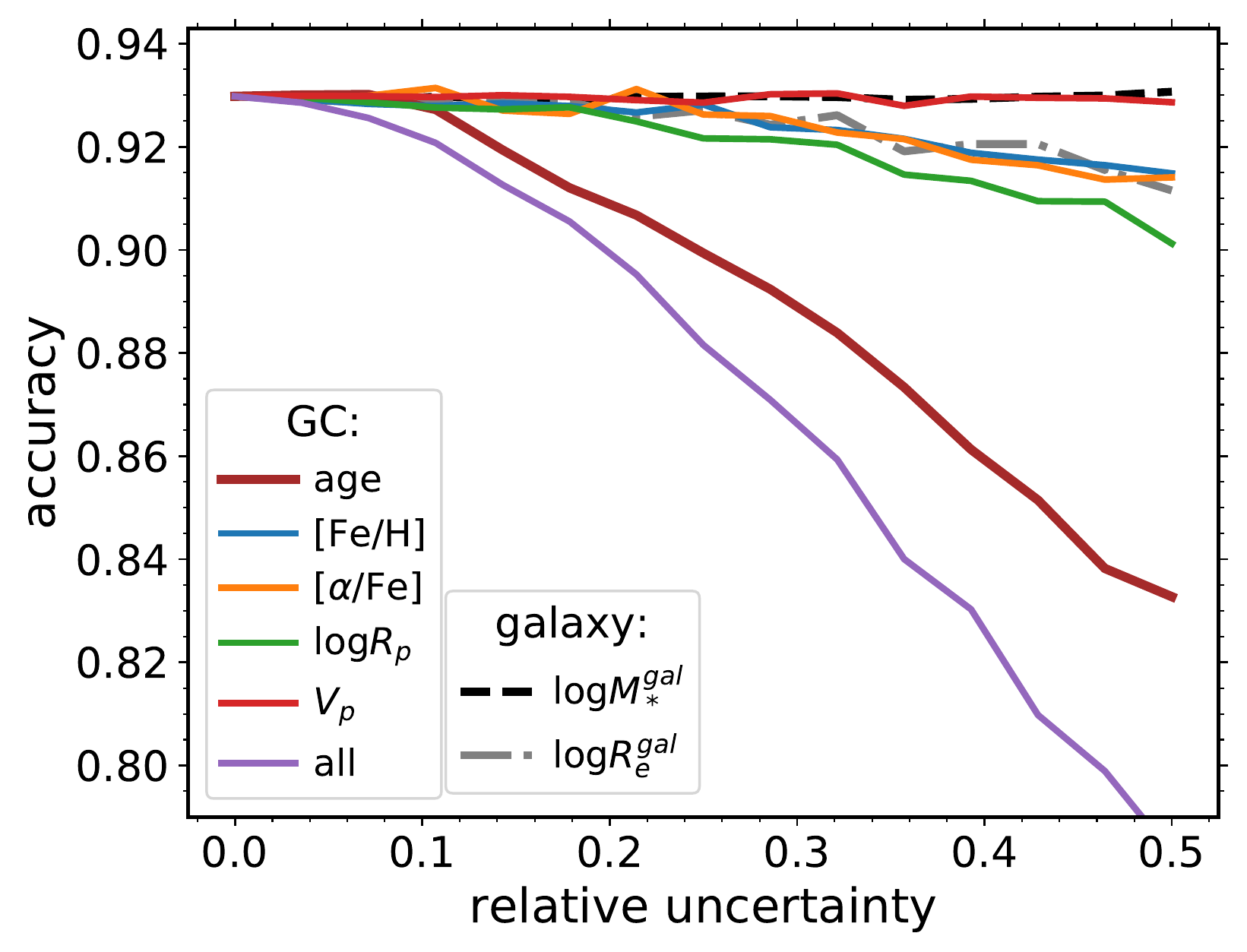}
    \caption{Accuracy of a ANN classifier that includes GC ages in addition to all the features of the fiducial model. Each line shows the accuracy as a function of the relative error in the GC observables: metallicity, alpha abundances, projected position, and line-of-sight velocity. As in Fig.~\ref{fig:accuracy_vs_obserrors}, the uncertainties in GC velocities are expressed as a fraction of the galaxy velocity dispersion, and for all other observables (including ages) in logarithmic units. The bottom line shows the combined effect of uncertainties in all the observables. The values are obtained by adding normally distributed Monte Carlo errors to the test set drawn from the simulations. Including GC ages significantly improves the accuracy of the predictions for uncertainties $\la 0.1$ dex, but makes the model very sensitive to the precision of the age measurements.}
\label{fig:accuracy_vs_obserrors_ages}
\end{figure}

The impact of including GC ages on the predictions is shown in Figure~\ref{fig:accuracy_vs_obserrors_ages}. Including ages increases the accuracy of the model with no uncertainties from $\sim 89$ to $\sim 93$ per cent. Relative uncertainties up to $\sim 0.2$ in the GC metallicities, alpha-abundances, positions and velocities have almost no effect on the accuracy in this model. However, the classifier performance drops significantly when the precision of the ages is reduced below $\sim 0.1$ dex. This demonstrates the importance of the GC ages compared to all the other observables in shaping the model predictions. For reference, recent advances in stellar population modelling make possible to achieve this level of precision in the age determination of clusters \citep[see][]{Cabrera-Ziri22}. For ages with a precision of $\la 0.1$ dex (or about 25 per cent), the classifier reaches an accuracy of $> 92$ per cent, suggesting that the current limiting precision of GC ages is already high enough to significantly improve the performance of the ANN model.

\section{Discussion}
\label{sec:discussion}

The GC observables we select in this work have been found to be good indicators of GC origin in previous studies. \citet{Hughes19b} found that the alpha-element abundance of recently accreted GCs is systematically lower at fixed [Fe/H] relative to in-situ GCs. \citet{emosaicsII, krakenI} showed that at a fixed age, the metallicity of GCs traces the metallicity of their galactic progenitor, and therefore GC age-metallicity relations can be used to reconstruct the assembly history of the Galaxy. \citet{Pfeffer20} and \citet{krakenII} showed that adding 3D orbital information to the ages and metallicities allows the recovery of the masses and accretion redshifts of each progenitor. 

We have extensively explored the space of GC and galaxy properties to use as features for reconstructing GC origin. In addition to our manual exploration, our choice of classifier model architecture (a densely layered neural network) is meant to take advantage of the ability of these networks to capture highly non-linear relationships between the features and the output. It is therefore unlikely that we excluded a feature in the simulations that would dramatically improve the performance of the classifier. More sophisticated simulations that track the individual abundances of many isotopes \citep[e.g.][]{Reina-Campos22} may capture additional information that could improve the predictions.

Another more subtle issue that arises in this type of machine learning problem is the completeness of the training set. Due to the steepness of the galaxy stellar mass function, our volume-limited simulated galaxy sample is dominated by low-mass galaxies and contains only a handful of massive elliptical galaxies. While this provides an unbiased representation of the galaxy population, it is not ideal for supervised learning. As shown in Section~\ref{sec:performance}, the classifier has difficulties capturing the relation between GC observables and their origin in the most massive galaxies partly due to the small size of the galaxy training sample, which only includes 4 galaxies with $\Mstar>10^{11}\Msun$ compared to 86 MW-mass ($10^{10}<\Mstar/\Msun<10^{11}$) galaxies and 273 dwarfs in the mass range of the Magellanic Clouds ($4\times10^8<\Mstar/\Msun<3\times10^9$). Similarly, our results indicate that the loss of information when the phase-space distribution of the GC systems is observed in projection is one of the dominant limiting factors in the performance of our model, compared to one that takes as input the full 6D information. We have explicitly tested this hypothesis using the `data augmentation' technique. This was done by retraining the model using an extended training set that includes three orthogonal projections of the simulation box, instead of the single projection used for the fiducial model. This procedure effectively yields a three times larger training set. There was no significant improvement in the predictive accuracy, which suggests that the method is limited only by the lack of depth information, and not by the number of galaxies (or projections) in the training set. A further limitation of the model presented here is that the selection of the training sample implies that the results may apply only to central galaxies. Achieving a similarly good performance on satellites will likely require training specifically with satellite galaxies because their evolution is more sensitive to environmental processes.

Lastly, there is the possibility that the simulation used for training the algorithm does not capture certain aspects of the formation and evolution of galaxies and GCs, and this is the most difficult aspect of the uncertainties to quantify. As described in Sec.~\ref{sec:simulations}, E-MOSAICS reproduces many properties of observed galaxies and GCs. However, the EAGLE model produces $L^*$ galaxies with stellar masses that are $\sim 0.1{-}0.2$ dex below observations \citep{Schaye15}. Furthermore, the lack of a cold interstellar medium in EAGLE results in the artificial survival of too many young, metal-rich clusters that should have otherwise disrupted (for a detailed discussion, see \citealt{emosaicsI} and \citealt{emosaicsII}). While the first problem is difficult to correct for in the training of the ANN, Fig.~\ref{fig:accuracy_vs_obserrors} suggests that the predictions are robust to large errors in the stellar mass. We have also attempted to remove the underdisrupted GCs from the training and test samples. A new generation of simulations with better modelling of $L^*$ galaxies and improved ISM physics will be needed to extend the origin predictions to metal-rich GCs with $\feh>-0.5$ \citep[see][]{Reina-Campos22}, and will likely improve the identification of in-situ objects (see Sec.~\ref{sec:performance}).

The reconstruction of the Milky Way assembly history using \emph{Gaia} and other spectroscopic surveys demonstrates that chemo-dynamical observations are a powerful tool. Thanks to these studies, the origin of the stellar halo of the MW has now been determined as a function of galactocentric radius \citep{Naidu20}. This and other detailed observations like the radial profile of galactic components of different origin could become excellent tools to constrain cosmological hydrodynamical simulations. Simulations have already reached enough sophistication to reproduce many global galaxy observables, but still suffer from highly degenerate input physics, which limits their predictive power \citep[for a review, see][]{NaabOstriker17}. The deep learning approach we demonstrate in this paper could in principle be extended to constrain the spatial distribution of in-situ and accreted stars and GCs in galaxy samples of up to millions of objects in the local Universe. Classifiers trained using observables that are independent of specific highly uncertain physical processes (i.e. stellar  feedback) could determine the spatial distribution of in-situ and accreted material across the galaxy population. By comparing these constraints to the output of state-of-the-art cosmological simulations, their built-in hypotheses regarding the physics of star formation and feedback could be tested. Similar methods could be employed to constrain the physics of the DM particle using galaxy surveys.

\section{Conclusions}
\label{sec:conclusions}

In this work we use nearly a thousand simulated galaxies and their GC systems in the E-MOSAICS (34.4 Mpc)$^3$ periodic volume to understand how the present day GC observables (e.g. metallicity, alpha abundances, projected distance and velocity) can be used to infer the origin of specific GCs (i.e. in-situ vs. accreted). We first investigate how galaxy properties including halo mass and metallicity influence the fraction of GCs that are accreted from satellites across the galaxy mass spectrum, from dwarfs to giant ellipticals. In the second part we use supervised deep learning algorithms to model and understand the relation between GC observables in external galaxies and their in-situ or accreted origin. For this we exploit the success of the E-MOSAICS cluster formation and evolution physics in reproducing the observed properties of GCs in the local Universe. We train a Multilayer Perceptron artificial neural network on the mapping between 17 GC and host galaxy observable features (see Table~\ref{tab:table2}), and their true origin labels (i.e. in-situ versus accreted). We test the performance of the classifier on an independent random subset comprised of $\sim 20$ per cent of the simulated galaxies, and use the known origin of the Milky Way GCs to benchmark the model for application on extragalactic GC systems. We investigate the importance of each observable for determining the predictions of the classifier, and the effect that uncertainties in the observations have on the accuracy of the predictions. Finally, we explore the benefits of including GC ages.  

Our conclusions are summarised as follows:
\renewcommand{\labelenumi}{\arabic{enumi}.}
\begin{enumerate}
    \item The balance of in-situ formation and accretion of GCs is strongly shaped by galaxy mass, in a similar way as for the field stars. The median accreted fraction of GCs increases with mass, such that dwarf galaxies are typically dominated by in-situ GCs, and massive ellipticals contain mostly accreted GCs (Fig.~\ref{fig:accretedfrac_Mhalo}). Despite the large scatter in accreted GC fractions across the simulated galaxies, we find a weak trend with halo mass: at fixed stellar mass, galaxies in more massive haloes host larger fractions of accreted GCs (Fig.~\ref{fig:accretedfrac_Mhalo}). Metal-poor galaxies also tend to have larger accreted GC fractions due to a larger contribution of relatively metal-poor satellites to their assembly, and the late formation of their DM haloes (Fig.~\ref{fig:accretedfrac_feh}).  
    \item There is a strong dependence of GC origin on GC metallicity. Metal-poor GCs are typically a mix of in-situ and accreted objects, whereas the origin of metal-rich GCs depends on stellar mass: in low-mass galaxies (with $\Mstar<10^{10}\Msun$) they are almost entirely formed in-situ, and in galaxies more massive than the Milky Way they are mostly accreted  (Figs.~\ref{fig:accretedfrac_GCmetbins} and \ref{fig:GC_origin_vs_feh}).
    \item A Multilayer Perceptron artificial neural network classifier trained on the observable properties of more than 50,000 GCs hosted by 736 simulated galaxies predicts the in-situ/accreted origin of GCs in a test sample drawn from the same simulation with an overall accuracy of $\sim 89$ per cent for objects with unambiguous labels (with a completeness of 60 per cent; Sec.~\ref{sec:training} and \ref{sec:performance}). The classifier is excellent at identifying accreted GCs (6 per cent false-positive rate), and less accurate for in-situ GCs (18 per cent false-positive rate) (Fig.~\ref{fig:confusion_matrix}). The model performs generally well in low-mass galaxies (below the mass of the MW), but has more difficulty identifying in-situ GCs in the most massive galaxies (Figs.~\ref{fig:accuracy_galprops} and \ref{fig:projections}). This is likely due to the similarity of the observables of in-situ and accreted populations in massive galaxies (Fig.~\ref{fig:corner_plots_ellipticals}), their low fraction of in-situ GCs, the small number of these galaxies in the simulated volume ($\sim 6$), and the exclusion of GCs with $\feh>-0.5$ from the sample.
    \item The classifier uses only a few dominant observables to predict GC origin. These include the effective radius, stellar mass, and alpha-element abundance of the host galaxy, together with the GC metallicity and alpha-abundance relative to the galaxy, and its projected angular momentum and galactocentric radius (Fig.~\ref{fig:feature_importance}). The high predictive importance of the galaxy effective radius seems to originate from its correlation with the assembly timescale of the galaxy and its effect on the GC accreted fraction (see Sec.~\ref{sec:properties}). Simulated galaxies with larger effective radii formed later and in more massive DM haloes with larger accreted fractions.
    \item Using the simulated test data, we find a significant correlation between the mean prediction confidence (an output of the ANN classifier) and the accuracy for each galaxy. This allows us to estimate the likelihood that predictions for GC origin in a real galaxy will reach a minimum desired accuracy (Fig.~\ref{fig:accuracy_confidence}). 
    \item After removing observables that are either unimportant or difficult to obtain, we test a minimal version of the classifier on the Milky Way GCs with known origin. Assuming that the Galaxy is observed in projection, the optimized model achieves excellent performance, with an accuracy of $\sim 85{-}90$ per cent that is nearly independent of the inclination. The model identifies GCs associated to each of the five known GC-rich progenitor galaxies, including most of the GCs accreted from \emph{Kraken}, and all of the \emph{Gaia-Enceladus} GCs.
    \item The classifier is robust to relatively large uncertainties in the observational data (i.e. larger than in currently available extragalactic data). Relative uncertainties in the GC metallicity, alpha-abundances, projected distance, and line-of-sight velocity of up to $\sim 0.1$ decrease the accuracy on the test data by less than 1 per cent. The dominant effect is from the uncertainty in the alpha-abundances and galaxy effective radii (Fig.~\ref{fig:accuracy_vs_obserrors}).
    \item Including GC ages as an additional feature in the model significantly increases the performance, with an accuracy on the simulation test data of $> 92$ per cent. However, a precision of $<0.1$ dex (or $\sim 25$ per cent) is required in the age measurements. Ages with lower than $0.1$ dex precision produce a steep decrease in performance compared to the fiducial model (Fig.~\ref{fig:accuracy_vs_obserrors_ages}).
  
\end{enumerate}

The ANN classifier developed in this work can be readily used to make predictions for the origin of GCs in nearby galaxies for which metallicity, alpha-abundance, positions, and radial velocities have been measured. Over the next decade, wide-field space-based surveys will allow these data to be collected for very large samples of galaxies. The model developed in this work is the initial step in piecing together the assembly histories of galaxies beyond the Milky Way as a function of mass and environment, leading to a detailed understanding of the process of galaxy formation. In future work we will explore efficient methods to constrain galaxy merger histories using GC observables. In a follow-up paper we apply the model to predict the origin of the GCs in M31 (Trujillo-Gomez et al. in prep.).

The python implementation of the fiducial and minimal classifiers in {\sc Keras}, along with an example of their use in an interactive {\sc Jupyter} notebook is available at \url{https://github.com/sebastian-tg/GC-origin-ANNclassifier}.


\section*{Acknowledgements}

STG gratefully aknowledges funding by the Deutsche Forschungsgemeinschaft (DFG, German Research Foundation) -- Project-ID 138713538 -- SFB 881 (``The Milky Way System'', subproject A08). STG and JMDK gratefully acknowledge funding from the European Research Council (ERC-StG-714907, MUSTANG). JMDK gratefully acknowledges funding from the German Research Foundation (DFG - Emmy Noether Research Group KR4801/1-1). COOL Research DAO is a Decentralised Autonomous Organisation supporting research in astrophysics aimed at uncovering our cosmic origins. MRC gratefully acknowledges the Canadian Institute for Theoretical Astrophysics (CITA) National Fellowship for partial support. JP is supported by the Australian government through the Australian Research Council's Discovery Projects funding scheme (DP200102574). RAC is supported by the Royal Society. NB gratefully acknowledges financial support from the European Research Council (ERC-CoG-646928, Multi-Pop) as well as from from the Royal Society (University Research Fellowship). This study was supported by the Klaus Tschira Foundation. This work used the DiRAC Data Centric system at Durham University, operated by the Institute for Computational Cosmology on behalf of the STFC DiRAC HPC Facility (\url{www.dirac.ac.uk}). This equipment was funded by BIS National E-infrastructure capital grant ST/K00042X/1, STFC capital grants ST/H008519/1 and ST/K00087X/1, STFC DiRAC Operations grant ST/K003267/1 and Durham University. DiRAC is part of the National E-Infrastructure. The work also made use of high performance computing facilities at Liverpool John Moores University, partly funded by the Royal Society and LJMU's Faculty of Engineering and Technology.

This work made use of the software packages: {\sc Numpy} \citep{numpy}, {\sc Scipy} \citep{scipy}, {\sc Matplotlib} \citep{matplotlib}, {\sc Pandas} \citep{pandas}, {\sc Seaborn} \citep{seaborn}, {\sc Jupyter} \citep{jupyter}, {\sc Pynbody} \citep{pynbody}, {\sc Scikit-learn} \citep{scikit-learn}, {\sc Tensorflow} \citep{tensorflow}, and {\sc Keras} \citep{keras}.

\section*{Data availability}

The data underlying this article will be made available upon reasonable request to the corresponding author.



\bibliographystyle{mnras}
\bibliography{merged}


\appendix

\section{Clustering analysis for feature importance}
\label{sec:clustering_analysis}

\begin{figure*}
    \includegraphics[width=0.95\textwidth]{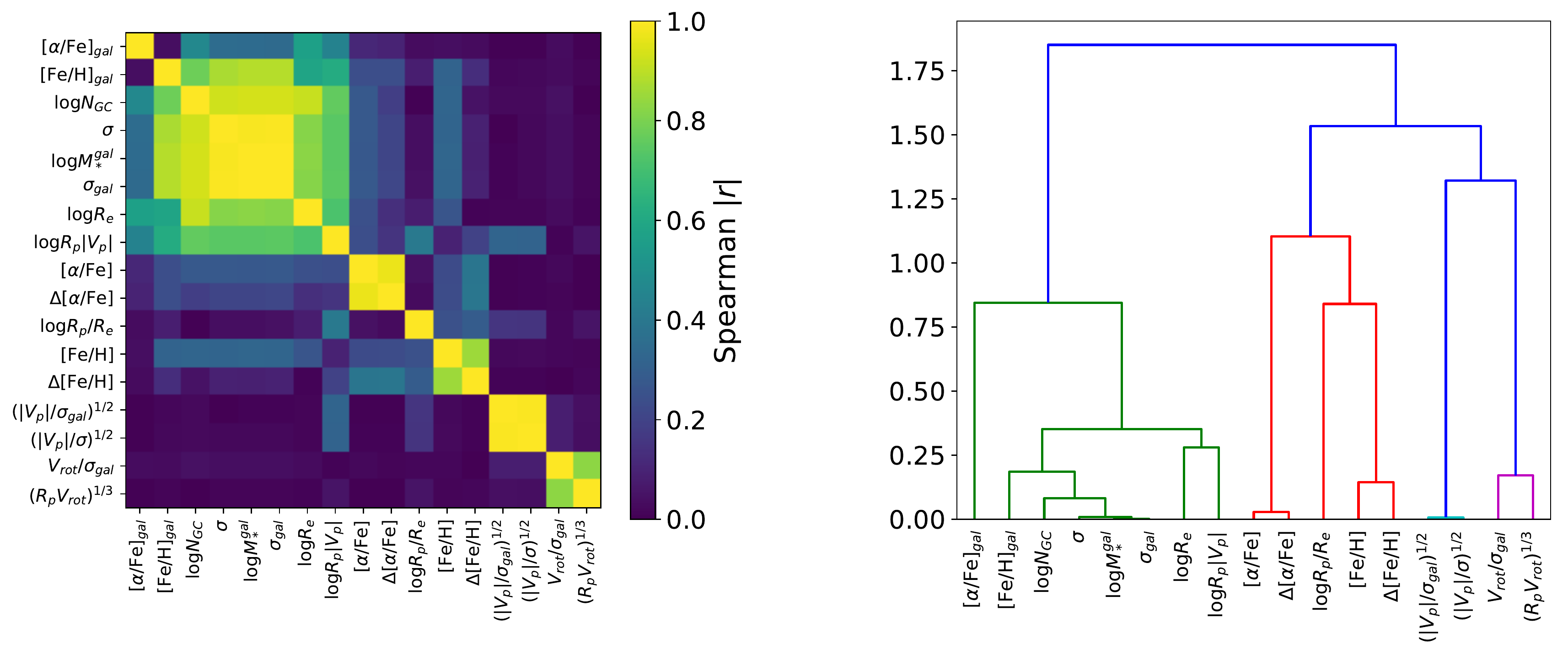}
    \caption{Clustering of GC and galaxy features. Left: Spearman correlation matrix. Right: dendrogram of correlated feature clusters.}
\label{fig:clustering}
\end{figure*}

Figure~\ref{fig:clustering} shows the results of the clustering analysis for the features of the fiducial ANN classifier in Sec.~\ref{sec:importances}. A threshold of 0.25 was used to select representative features in each covariant group. A new classifier was then trained on the selected subset of independent features to obtain an unbiased estimate of the predictive importance of each one.

\section{Distribution of GC observables with the highest predictive power}
\label{sec:corner_plots}

\begin{figure*}
    \includegraphics[width=0.80\textwidth]{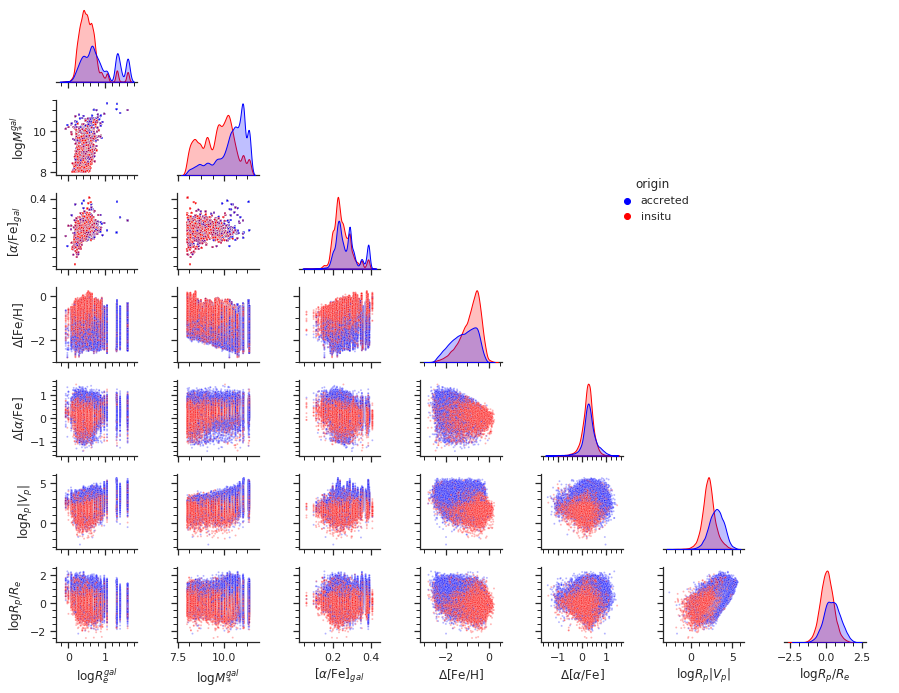}
    \caption{Joint and marginal distributions of GC origin across the galaxy and GC observables with the most predictive power. The panels show the distribution of in-situ and accreted GCs across the entire simulated sample in the space of the observables with the most predictive power (see Fig.~\ref{fig:feature_importance} and Sect.~\ref{sec:importances}). In-situ GCs are coloured red, while accreted GCs are shown in blue. Significant overlap of the two classes limits the predictive power of individual observables, but the ANN classifier is able to combine them optimally.}
\label{fig:corner_plots}
\end{figure*}

Figure~\ref{fig:corner_plots} shows the joint distribution of the seven most important GC origin predictors across the entire simulated GC sample, with colour indicating their true origin. The observables with the highest importance also show the most distinct separation in the distributions of in-situ and accreted GCs. This qualitatively confirms the result of the permutation importance analysis, and shows that the classifier effectively uses the GC and galaxy properties that correlate most with GC origin.

\section{Performance of the minimal classifier}
\label{sec:minimal_model}

\begin{figure}
    \includegraphics[width=\columnwidth]{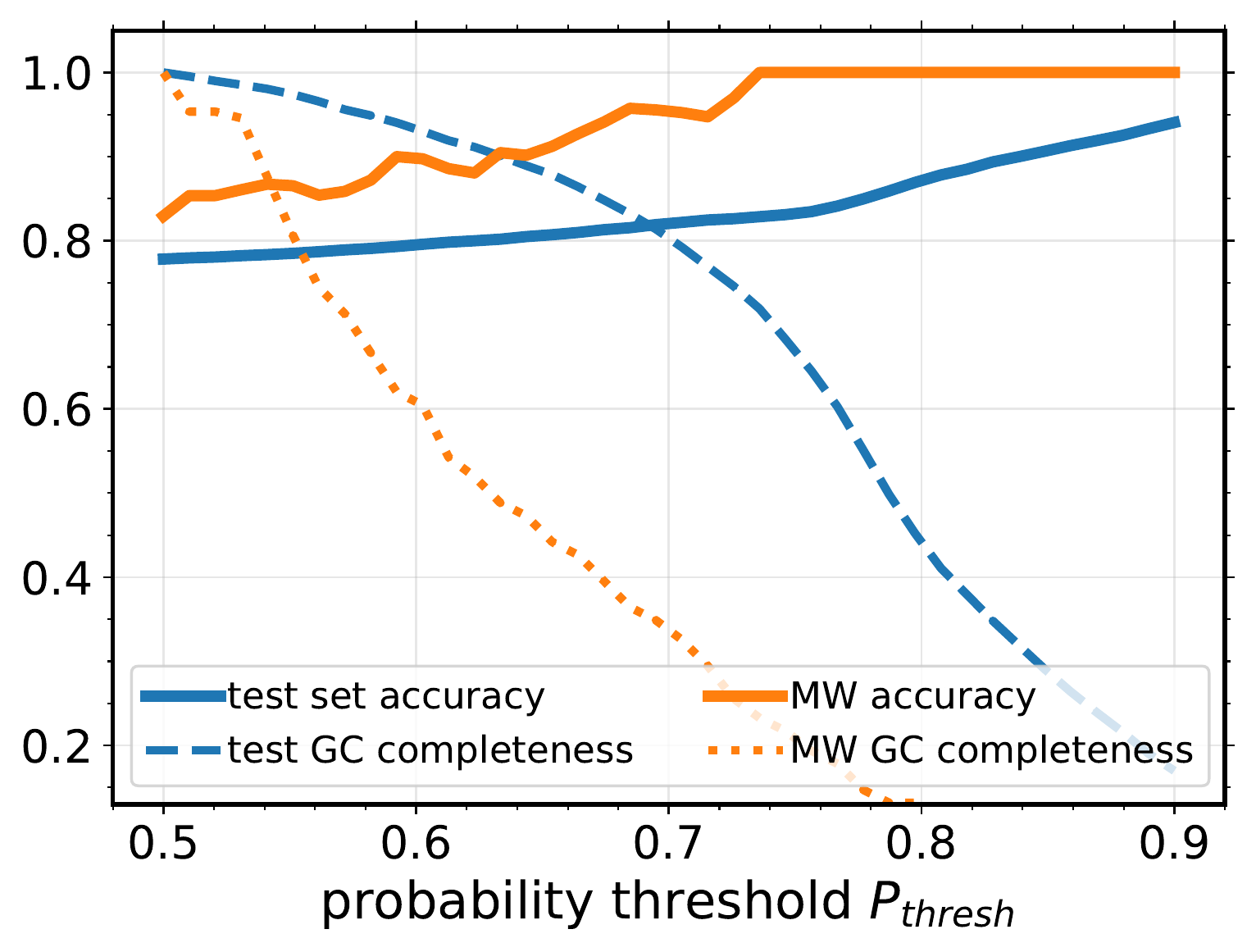}
    \caption{Accuracy and sample completeness of predictions as a function of the decision threshold adopted for the minimal classifier. The coloured solid lines indicate the performance on the simulation test set (blue), and on the Milky Way GC system (orange).}
\label{fig:accuracy_threshold_MWtest}
\end{figure}

Figure~\ref{fig:accuracy_threshold_MWtest} shows the performance of the best minimal classifier and the fraction of unambiguous predictions as a function of the adopted decision threshold. The accuracy is shown for both the simulation GC test set and the MW GC system. These values can be used to obtain a rough estimate of the expected performance on observed galaxies.


\bsp	
\label{lastpage}
\end{document}